\documentclass[12pt]{article}
\usepackage{latexsym,amssymb, amsmath,color,graphicx,fullpage,comment,cancel,youngtab,slashed,bm}
\usepackage[normalem]{ulem}
\usepackage{epstopdf,tabularx}
\usepackage[utf8]{inputenc}
\usepackage{authblk,array}
\usepackage[hidelinks]{hyperref} 
{ 
\date{\vspace{-5ex}}
\begin{document}
\begin{titlepage}
\begin{flushright}\small{MCTP-14-34} 
\end{flushright}

\begin{center}
\vspace{1cm}

{\Large \bf Simultaneous B and L Violation: \\ New Signatures from RPV-SUSY}

\vspace{0.8cm}

\small
\bf{Cyrus Faroughy$^{1}$,  Siddharth Prabhu$^{2}$, Bob Zheng$^{3}$}
\normalsize

\vspace{.5cm}
{\it $^1$ Department of Physics and Astronomy, Johns Hopkins University, Baltimore, MD 21218}\\
{\it $^2$ Department of Physics, Yale University, New Haven, CT 06511} \\
{\it $^3$ Michigan Center for Theoretical Physics, University of Michigan, Ann Arbor, MI 48109} \\

\end{center}

\vspace{1cm}

\begin{abstract}
\noindent Studies of R-parity violating (RPV) supersymmetry typically assume that nucleon stability is protected by approximate baryon number (B) or lepton number (L) conservation. We present a new class of RPV models that violate B and L simultaneously (BLRPV), without inducing rapid nucleon decay. These models feature an approximate $Z_2^e \times Z_2^\mu \times Z_2^\tau$ flavor symmetry, which forbids 2-body nucleon decay and ensures that flavor antisymmetric $L L E^c$ couplings are the only non-negligible L-violating operators. Nucleons are predicted to decay through $N \rightarrow K e \mu \nu$ and $n \rightarrow e \mu \nu$; the resulting bounds on RPV couplings are rather mild. Novel collider phenomenology arises because the superpartners can decay through both L-violating and B-violating couplings. This can lead to, for example, final states with high jet multiplicity and multiple leptons of different flavor, or a spectrum in which depending on the superpartner, either B or L violating decays dominate. BLRPV can also provide a natural setting for displaced $\tilde{\nu} \rightarrow \mu e$ decays, which evade many existing collider searches for RPV supersymmetry. 

\end{abstract}
\end{titlepage}
%\tableofcontents
\section{Introduction}

R-parity violating (RPV) \cite{Aulakh:1982yn,Hall:1983id,Ross:1984yg,Ellis:1984gi,Dawson:1985vr,Barbier:2004ez} supersymmetry (SUSY) has become increasingly well motivated, due to comparatively weaker collider bounds on colored sparticle production \cite{Carpenter:2006hs,Brust:2011tb}, along with null results for MSSM dark matter. If both baryon number (B) and lepton number (L) violating RPV couplings are present, 4-fermion effective operators of the form $q q q \ell$ will induce 2-body nucleon decay \cite{Hinchliffe:1992ad}. Consequently, Super-Kamionkade bounds on 2-body nucleon decay ($\tau_{N \rightarrow M \ell} \gtrsim 10^{34}$ years) \cite{Nishino:2012ipa} strongly constrain the product of B and L violating couplings for $\sim$ TeV scale superpartners. In order to avoid these bounds, the canonical approach is to assume that either B or L is approximately conserved. This leads most authors to consider two broad classes of RPV SUSY models: those which violate B (BRPV), and those which violate L (LRPV).

The main purpose of this paper is to establish the existence of a third class of RPV models: those which violate B and L \emph{simultaneously}. We refer to this class of models using the acronym BLRPV\footnote{Simultaneous B and L violation was recently discussed in \cite{An:2013axa} as a potential baryogenesis  mechanism.}. This possibility arises because $\lambda_{i j k} L_i L_j E^c_k$ superpotential couplings which are antisymmetric in flavor indices\footnote{To be precise, they are antisymmetric in flavor indices in the mass eigenstate basis of charged leptons.} i.e. $i \neq k, j \neq k$ do \emph{not} generate dangerous 4-fermion $q q q \ell$ effective operators in the presence of the $\lambda^{\prime \prime} U^c D^c D^c$ BRPV couplings. 

Instead, the combination of $\lambda^{\prime \prime} U^c D^c D^c$ and $\lambda_{i j k} L L E^c,\, i \neq k, j \neq k$ couplings generate 6-fermion effective operators, resulting in the nucleon decay modes $N \rightarrow K \nu e^\pm \mu^\mp$ and $n \rightarrow e^\pm \mu^\mp \nu$. There have been no recent experimental attempts to search for these decay modes; a discovery in these channels would provide strong evidence for BLRPV. We will show that the experimental constraint $\tau\left(N\rightarrow \mu^+ \,\mathrm{inclusive}\right) \gtrsim 10^{32}$ years \cite{Cherry:1981uq} results in the bound $\left|\lambda^{\prime \prime}_{ 1 1 2} \lambda_{i j k} \right| \lesssim 10^{-10}$ for $i \neq k, j \neq k$ assuming $\sim 1$ TeV superpartners; this bound weakens to $\sim 10^{-4}- 10^ {-3}$ for $\lambda^{\prime \prime}$ couplings with heavy flavors. Such comparatively weak bounds allow both $\lambda^{\prime \prime} U^c D^c D^c$ and $\lambda_{i j k} L_i L_j E^c_k, \, i \neq k, j \neq k$ couplings to be relevant for collider phenomenolgy, without violating nucleon decay bounds. Although we focus here on RPV-SUSY, our computation of nucleon decay rates from 6-fermion effective operators is of general interest for BSM theories with B and L violation. These computations have not appeared elsewhere in the literature.

For a consistent model of BLRPV, one expects a symmetry to enforce the flavor antisymmetry of $ L_i L_j E^c_k$ while suppressing/forbidding all other LRPV couplings. If $\lambda_{i j k } L_i L_j E^c_k, \, i \neq k, j \neq k$ are the only non-vanishing LRPV couplings, there is a $Z_2^e \times Z_2^\mu \times Z_2^\tau$ flavor symmetry which is exact in the absence of neutrino masses. Under $Z_2^\tau$, both $L_\tau$ and $E^c_\tau$ are even while all other lepton superfields are odd; $Z_2^\mu$ and $Z_2^e$ are similarly defined. This symmetry (which is anomaly-free in the sense of \cite{Krauss:1988zc,Ibanez:1991pr}) forbids all effective operators of the form $q q q \ell$, providing an intuitive explanation for the absence of 2-body proton decay. $Z_2^e \times Z_2^\mu \times Z_2^\tau$ must be broken by neutrino masses and mixing angles for realistic neutrino phenomenology \cite{Leurer:1992wg}; we will argue that the resulting bounds from $q q q \ell$ operators induced by neutrino masses are mild. 

The collider phenomenology of BLRPV can give distinct LHC signatures which differentiates it from standard RPV SUSY. Such novel phenomenology occurs because in BLRPV, sparticles can decay via both LRPV and BRPV couplings. For example, if a sparticle's decay rates via BRPV and LRPV are comparable, new signatures from sparticle pair production arise e.g. $\tilde{q} \tilde{q} \rightarrow q q \chi^+ \chi^- \rightarrow 5q \, e \mu \tau$. Such final states will be characterized by high jet multiplicity, three hard leptons of different flavor, and no missing energy. Alternatively, different sparticles can decay predominantly via either BRPV or LRPV. For instance, a mostly RH squark can decay predominantly via the $U^c D^c D^c$ coupling, while the LH squark decays to the LSP which subsequently decays through  $L L E^c$. We discuss illustrative examples which highlight these qualitative features, saving a more general study for future work. 

Finally, we discuss the phenomenology of displaced $\tilde{\nu} \rightarrow \mu e$ decays in BLRPV-SUSY. Though this decay mode is certainly not unique to BLRPV, it can occur naturally in the BLRPV framework for a $\tau$-sneutrino LSP, due to the flavor structure in $L L E^c$ couplings enforced by $Z_2^e \times Z_2^\mu \times Z_2^\tau$. It has been noted in \cite{Batell:2013bka} that displaced $\tilde{\nu} \rightarrow \mu e$ decays with $c \tau \sim \mathcal{O}( \mathrm{cm})$ can mitigate collider constraints on superpartners decaying via LRPV. In the BLRPV framework, the $\tau$-sneutrino also has competing 4-body decay modes $\tilde{\nu}_\tau \rightarrow \tau \overline{d}_{i} \overline{d}_j \overline{d}_k$ and $\tilde{\nu}_\tau \rightarrow \nu u_i d_j d_k$ which occur via the BRPV coupling $\lambda^{\prime \prime}_{i j k}$. Focusing on the (least constrained) $\lambda^{\prime \prime}_{3 3 2}$ coupling, we compute $\Gamma(\tilde{\nu_\tau} \rightarrow \tau\, \overline{s} \,\overline{b} \, \overline{b})$ and show that the BRPV branching ratio is small for large regions of parameter space, even if $c \tau \sim \mathcal{O}(\, \mathrm{cm})$ and $\left|\lambda^{\prime \prime}_{3 3 2}\right| \gtrsim 0.1$. It is therefore possible to have spectra where e.g. the stop decays via BRPV while the $\tau$-sneutrino has displaced decay to $\mu e$.

This paper is organized as follows. In Section \ref{constraints}, we compute nucleon decay bounds on $\left| \lambda^{\prime \prime} \lambda_{i j k}\right|$ for $i \neq k, j \neq k$. In Section \ref{flavorsym}, we discuss the $Z_2^e \times Z_2^\mu \times Z_2^\tau$ symmetry of BLRPV, and discuss implications of $Z_2^e \times Z_2^\mu \times Z_2^\tau$ breaking due to neutrino masses. In Section \ref{collider}, we discuss the collider phenomenology of BLRPV. In Section \ref{displaced}, we discuss displaced $\tilde{\nu}_{\tau} \rightarrow \mu e$ decays in the BLRPV framework. We conclude in Section \ref{conclusion}. The appendices contain technical results which are referred to throughout the text.

\section{Nucleon Decay Phenomenology of BLRPV}\label{constraints}

The R-parity violating superpotential is given by\footnote{SM gauge symmetries enforce $\lambda^{\prime \prime}_{i j k} = -\lambda^{\prime \prime}_{i k j}$ and $\lambda_{i j k} = - \lambda_{j i k}$.}:\begin{equation}\label{WRPV} W_{RPV} = \frac{1}{2} \lambda^{\prime \prime}_{i j k} U^c_i D^c_j D^c_k + \lambda^{\prime}_{i j k} Q_i L_j D^c_k + \frac{1}{2} \lambda_{i j k} L_i L_j E^c_k + \kappa_i L_i H_u.\end{equation} To simplify terminology, we will refer to the $U^c D^c D^c$ operator as the BRPV operator, and the other lepton number violating operators as LRPV operators. 

\begin{figure}[t!]
\center{
\includegraphics[scale=1.1]{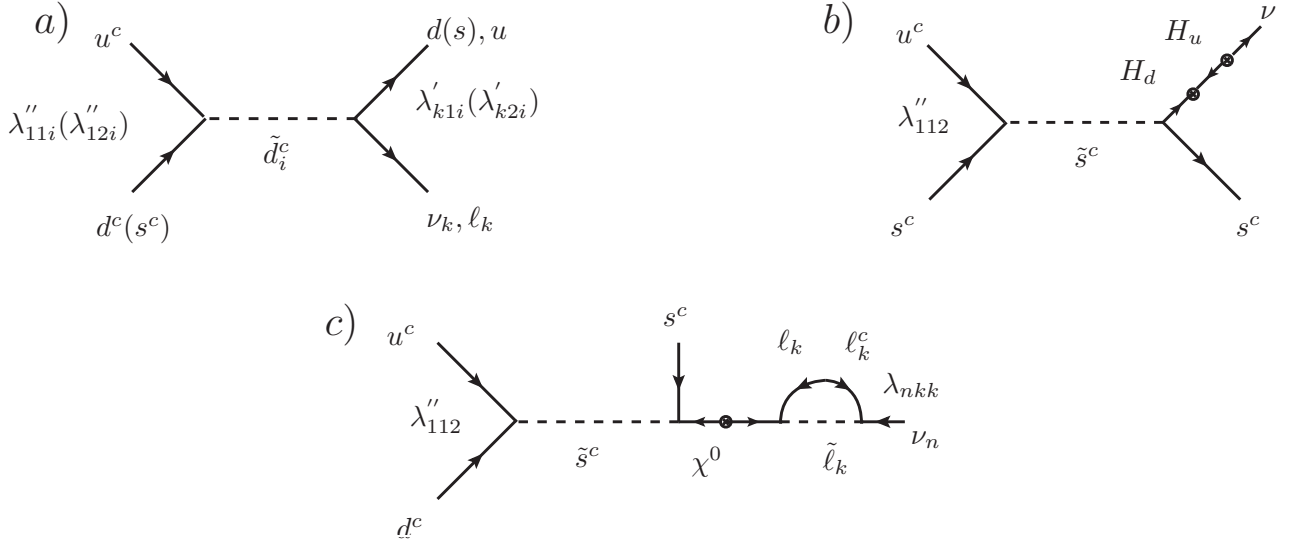}
 \caption{Diagrams which generate 4-fermion $q q q \ell$ operators  in the presence of BRPV and certain LRPV couplings. Such operators induce the 2-body nucleon decay modes $N \rightarrow M \ell$. \label{Fig:2body}}}
\end{figure}

If $\lambda^{\prime \prime}$ is non-vanishing, the presence of non-vanishing $\lambda^{\prime}$, $\lambda_{n k k}$ or $\kappa_i$ will induce 2-body nucleon decay modes such as $p\rightarrow \pi^+ \nu$ and  $p \rightarrow K^+ \nu$ via the 4-fermion effective operators depicted in Figure \ref{Fig:2body}.
Assuming a common superpartner mass scale $M_{SUSY}$, the Super-Kamiokande constraint $\tau_{N \rightarrow M \overline{\ell}} \gtrsim 10^{34}$ years \cite{Nishino:2012ipa} results in the bounds:
\begin{align}\label{2bodyconstraint}
\notag & \left|\lambda^{\prime \prime}_{1 1 k} \lambda^\prime_{1 j k}\right| \lesssim 10^{-25}\left(\frac{M_{SUSY}}{\mathrm{TeV}}\right)^2\text{\cite{Hinchliffe:1992ad}},\hspace{4mm}\, \left|\lambda^{\prime \prime}_{1 1 2} \frac{\kappa_i}{\mu}\right| \lesssim 10^{-23} \left(\frac{10}{\tan\beta}\right)\left(\frac{M_{SUSY}}{\mathrm{TeV}}\right)^2 \cite{Bhattacharyya:1998dt},\\ &\hspace{30mm}\left|\lambda^{\prime \prime}_{1 1 2} \lambda_{i j j}\right| \lesssim 10^{-23} \left(\frac{M_{SUSY}}{m_{\ell_j}}\right)\left(\frac{M_{SUSY}}{\mathrm{TeV}}\right)^2\text{\cite{Bhattacharyya:1998bx}}
\end{align} where $m_{\ell_j}$ is a SM lepton mass e.g. $m_{\ell_3} = m_\tau$. 

\begin{figure}[t!]
\center{
\includegraphics[width = 5 in]{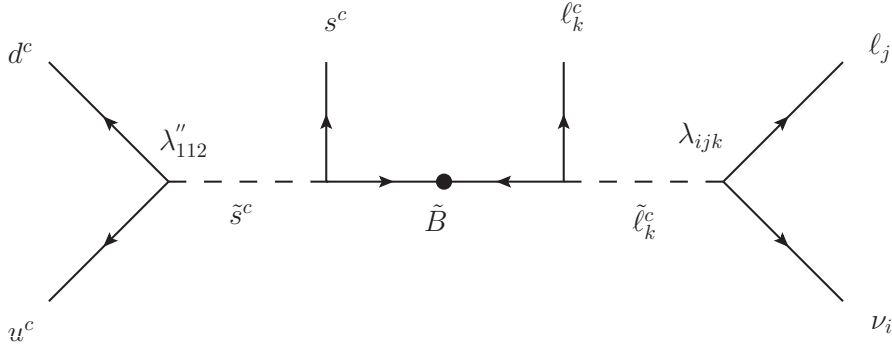}
 \caption{An example of the diagrams which generate 6-fermion effective operators in the presence of BRPV and $\lambda_{i j k},\, i \neq k, j \neq k$ couplings. Such operators induce the 4-body nucleon decay modes $N \rightarrow K \nu_i \ell^-_j \ell^+_k$.\label{Fig:4body}}}
\end{figure}

In RPV scenarios, the canonical approach is to satisfy (\ref{2bodyconstraint}) by assuming that either the BRPV or LRPV violating couplings are negligible. However, if the only non-vanishing LRPV couplings are $\lambda_{i j k} L_i L_j E^c_k $ for $i \neq k, j \neq k$ in the lepton mass eigenstate basis, the diagrams of Figure \ref{Fig:2body} vanish (absent flavor-changing slepton mass insertions). In this limit, the leading diagrams which induce nucleon decay are those similar to Figure \ref{Fig:4body}, resulting in 6-fermion effective operators and the 4-body decay modes $p \rightarrow K^+ \nu_i \ell^+_j \ell^-_k$, $n \rightarrow K \nu_i \ell^+_j \ell^-_k$. 

 In Section \ref{lightflavorconstraints}, we will compute $\Gamma(N \rightarrow K \nu_i \ell^+_j \ell^-_k)$ using chiral Lagrangian techniques, and obtain the experimental bound $\left| \lambda^{\prime \prime}_{1 1 2} \lambda_{i j k } \right|\lesssim 10^{-10}$ for $i \neq k, j \neq k$, assuming a common superpartner mass scale of 1 TeV. In Section \ref{heavyflavorconstraints}, we compute analogous bounds for $\lambda^{\prime \prime}_{i j k }$ couplings with at least 2 heavy ($c, b, t$) quark flavors, which induce nucleon decay via penguin-like loop diagrams with flavor changing $W^{\pm}, H^{\pm}, \chi^{\pm}$ exchange \cite{Smirnov:1996bg}. The resulting bounds are a factor of $10^{6}-10^7$ weaker than the corresponding bounds on $\lambda^{\prime \prime}_{1 1 2}$.

\subsection{Constraints on $\left|\lambda^{\prime \prime}_{1 1 2} \lambda_{i j k}\right|$, $i \neq k, j \neq k$ from $N \rightarrow K \nu_i \ell^+_j \ell^-_k$}\label{lightflavorconstraints}

A representative example of the diagrams relevant for nucleon decay in the presence of $\lambda^{\prime \prime}$ and $\lambda_{i j k }, i \neq k, j \neq k$ couplings is depicted in Figure \ref{Fig:4body}. There are also analogous diagrams involving virtual $\tilde{d}^c$, $\tilde{u}^c$, $\tilde{\ell}$ and $\tilde{\ell}^c$ exchange, corresponding to different permutations of the external quark and lepton legs. Taking into account all the relevant diagrams, integrating out sfermions in the limit of vanishing LR mixing leads to the following 6-fermion effective operators: \begin{align}\label{6fermion}
\mathcal{L}_{6f} = \notag &\left(\frac{2 \lambda^{\prime \prime}_{1 1 2} \lambda_{i j k} \,g_{Y}^2}{3 M_{\tilde{B}} \,\tilde{m}_{s_R}^2}\right)\Bigg[ \left(1+ 2 \frac{\tilde{m}_{s_R}^2}{\tilde{m}_{u_R}^2}\right) \Big(A_1 (u^c d^c) (s^c \ell^c_k) (\nu_i \ell_j)+ A_2 (u^c d^c) (s^c \ell_j) (\nu_i \ell^c_k)\Big) \\ &+ \left( \frac{\tilde{m}_{s_R}^2}{\tilde{m}_{d_R}^2} + 2 \frac{\tilde{m}_{s_R}^2}{\tilde{m}_{u_R}^2}\right) \Big(A_1 (u^c s^c) (d^c \ell^c_k) (\nu_i \ell_j)+ A_2 (u^c s^c) (d^c \ell_j) (\nu_i \ell^c_k)\Big)\Bigg]
\end{align} where $A_1 =  1/(2 \tilde{m}_{\nu_i}^2) + 1/\tilde{m}_{{\ell_R}_k}^2$ and $A_2 = 1/(2 \tilde{m}_{\nu_i}^2)-1/(2 \tilde{m}_{{\ell_L}_j}^2)$. The couplings which enter into (\ref{6fermion}) are evaluated at a renormalization scale near the superpartner mass scale. 

These $\Delta S = 1$ effective operators will induce the decay modes $p \rightarrow K^+ \nu_i \ell^+_j \ell^-_k$, $n \rightarrow K^0 \nu_i \ell^+_j \ell^-_k$. The $\lambda^{\prime \prime}_{1 1 2}$, $\lambda_{i j k }$ couplings also generate operators such as $(u^c d^c)(u^c \nu_j)(\nu_i \ell^c_k)$ and $(u^c d^c)(d^c \ell_j)(\nu_i \ell^c_k)$ via chargino exchange, which induce the 3-body decay modes $p \rightarrow \nu_i \nu_j \ell^+_k$ and $n \rightarrow \nu_i \ell^-_j \ell^+_k$ . The coefficients of these operators are flavor suppressed, such that the 3-body decays provide subdominant bounds if the squark flavor-changing mass insertions are not large. This is discussed in more detail in Appendix \ref{3bodydecay}.

In order to compute rates for nucleon decay, we match the effective operators in (\ref{6fermion}) to operators in the chiral Lagrangian \cite{Claudson:1981gh}. To do this, note that the operators in (\ref{6fermion}) transform as elements in the ${\bf  (1, 8)}$ representation of $SU(3)_L \times SU(3)_R$. To simplify the calculation, we assume slepton mass degeneracy and set $A_2 = 0$  in (\ref{6fermion}). Adopting 2-component spinor notation \cite{Dreiner:2008tw}, the corresponding operators in the chiral Lagrangian are \cite{Claudson:1981gh,Chadha:1983sj,Aoki:1999tw}:\begin{equation}\label{chirallagrangian}
\mathcal{L} = C_1\,\beta \left({\ell^c_k}\, Tr\left[\tilde{\mathcal{F}}^{\prime \prime} \xi^\dagger {B^c} \xi\right]\right)\left({\nu_i} {\ell_j}\right) + C_2\,\beta \left({\ell^c_k}\, Tr \left[\tilde{\mathcal{F}}^{\prime} \xi^\dagger {B^c} \xi\right]\right)\left({\nu_i} {\ell_j}\right)
\end{equation} where $\Psi_B = (B, {B^c}^\dagger)$ is the Dirac fermion corresponding to the baryon octet, $\xi = \exp{(i M/f_\pi)}$ where $M$ are the meson fields, and $\tilde{\mathcal{F}}^{\prime}, \tilde{\mathcal{F}}^{\prime \prime}$ are flavor projection matrices defined in \cite{Aoki:1999tw}. Parenthesis denote contraction of spinor indices. $\beta$ is related to the 3-quark annihilation hadronic matrix element i.e. $\left<0|(u_R d_R) u_R  |p \right> \approx \beta P_R u$ where $u$ is the spinor associated with the proton in Dirac notation. Lattice calculations give $\beta \approx 0.0118\, \left(\mathrm{GeV}\right)^3$ \cite{Aoki:2006ib}.

 \begin{figure}[t!]
\center{
\includegraphics[width = 6.0 in]{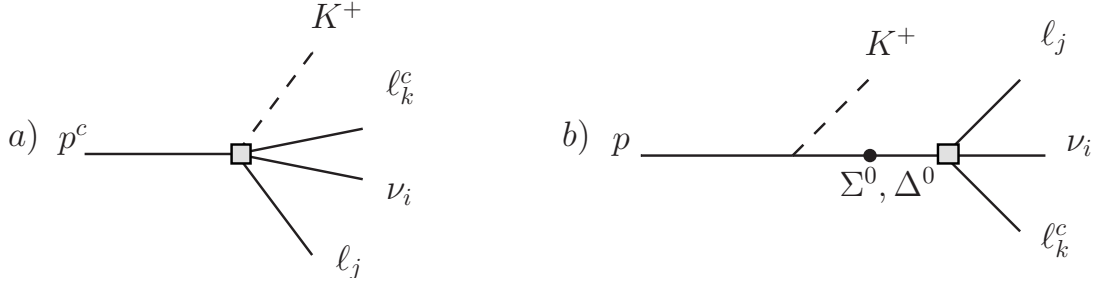}
 \caption{Diagrams which contribute to $p \rightarrow K^+\nu_i \ell^-_j \ell^+_k$. The shaded box represents the B and L violating vertex in the chiral Lagrangian.\label{Fig:protondecay}}}
\end{figure}

Matching the operators in (\ref{6fermion}) and (\ref{chirallagrangian}) fixes $C_1 = C_2 = 3 A\,\lambda^{\prime \prime}_{1 1 2} \lambda_{i j k}\,g_{Y}^2/M_{\tilde{B}}\,{\tilde{m}}^4$, where $\tilde{m}$ is the degenerate sfermion mass and $A \approx 0.22$ accounts for QCD effects which renormalize the effective operators in (\ref{6fermion}) from $Q = M_{SUSY}$ to $Q = \Lambda_{QCD}$ \cite{Ellis:1981tv}. Expanding the chiral Lagrangian along with the terms in (\ref{chirallagrangian}) to first order in $1/f_\pi$ then gives the necessary terms to compute nucleon decay amplitudes at tree level. This procedure was explicitly demonstrated in  \cite{Claudson:1981gh,Chadha:1983sj} and carried to completion in \cite{Aoki:1999tw} so we will not reproduce it here, though we have independently verified the relevant results. Although \cite{Aoki:1999tw} focuses exclusively on 2-body nucleon decay, the only difference between the present case and the amplitudes computed in \cite{Aoki:1999tw} is the addition of 2 leptons $\nu_i^\dagger \ell_j^\dagger$ \big(see (\ref{chirallagrangian})\big) at each $B$ and $L$ violating vertex.

Neglecting lepton masses, the $p \rightarrow K^+ \nu_i \ell^+_j \ell^-_k$ and $n \rightarrow K^0 \nu_i \ell^+_j \ell^-_k$ decay rates are given by\footnote{The amplitudes for $p \rightarrow K^+ \nu_i \ell^+_j \ell^-_k$ and $n \rightarrow K^0 \nu_i \ell^+_j \ell^-_k$ are related by approximate isospin invariance.}:\begin{equation}\label{4-bodydecay}
 \Gamma = \frac{\left|C_N\right|^2}{8192 \pi^5 M_N^3}\int_{{M_K}^2}^{{M_N}^2} \hspace{-4mm}  ds_{2 3 4} \int_0^{\left(\sqrt{s_{2 3 4}}- M_K\right)^2}\hspace{-4mm} ds_{2 3} \left(\frac{M_N^2 }{s_{2 3 4}}- 1\right)\left(M_N^2- s_{234}\right) s_{2 3} \lambda(s_{2 3 4}, s_{2 3}, M_K^2), \end{equation} where $\lambda(x, y, z) \equiv \left(x^2 + y^2 + z^2 - 2 x y - 2 x z - 2 z y\right)^{1/2}$ and \begin{equation}\label{coefficient}C_N = \frac{3 A \beta}{f_\pi} \left(\frac{\lambda^{\prime \prime}_{1 1 2} \lambda_{i j k} \,g_{Y}^2}{M_{\tilde{B}} {\tilde{m}}^4}\right) \left(1 + \frac{\left(D + 3 F\right) M_N}{3 M_\Lambda} + \frac{\left(D-F\right) M_N}{2  M_\Sigma}  + \frac{\left(D + 3 F\right) M_N}{6  M_\Lambda} \right).\end{equation} The baryon masses $M_\Sigma$, $M_\Lambda$ enter in (\ref{coefficient}) via diagrams with virtual $\Sigma$, $\Lambda$ exchange as shown schematically in Figure \ref{Fig:protondecay}; we neglect chiral symmetry breaking terms in the chiral Lagrangian.

In computing (\ref{4-bodydecay}) and (\ref{coefficient}), we have neglected terms in the amplitude proportional to $q^2= (p - k)^2$, where $p$ and $k$ are respectively the nucleon and kaon 4-momenta. This approximation is justified as $(p - k)^2 < (M_p - M_K)^2 \approx M_{p}^2/4$; a more precise calculation can be performed by including the $q^2$ terms which are given in \cite{Aoki:1999tw}. We use measured values for the chiral Lagrangian parameters $D = 0.8$ and $F=0.47$.

The effective operators in (\ref{6fermion}), (\ref{chirallagrangian}) will induce the decay modes $N \rightarrow K \nu e^\pm \mu^\mp$ where $N = p, n$.  There are currently no direct bounds on these decay modes (see e.g. \cite{Beringer:1900zz}). The strongest constraint on this process comes from the experimental constraint $\tau(N \rightarrow \mu^+ + \, \mathrm{anything}) \lesssim 10^{32}$ years \cite{Cherry:1981uq}, which imposes the bound:\begin{equation}\label{4bodybound}
\left|\lambda^{\prime \prime}_{1 1 2} \lambda_{i j k} \right| \lesssim 1 \times 10^{-10} \left(\frac{\overline{M}_{SUSY}}{\mathrm{TeV}}\right)^5
\end{equation} where $\overline{M}_{SUSY} \equiv ({\tilde{m}}^4 M_{\tilde{B}})^{1/5}$. For comparison, assuming a common superpartner mass $M_{SUSY}$,  bounds from dinucleon decay give $\left|\lambda^{\prime \prime}_{1 1 2} \right| \lesssim 10^{-6} (M_{SUSY}/\mathrm{TeV})^{5/2}$\,\cite{Goity:1994dq}, while bounds from loop-induced neutrino masses give $\left|\lambda_{1 2 3 } \lambda_{1 3 2 } \right| \lesssim 10^{-6}(M_{SUSY}/\mathrm{TeV})$\cite{Hall:1983id,Barbier:2004ez}\footnote{There are also subdominant bounds from flavor-changing processes, see e.g. \cite{Dreiner:2012mx,Dreiner:2013jta}.}. We reiterate that (\ref{4bodybound}) is only relevant if $i \neq k, j \neq k$, as all other $\lambda_{i j k}$ couplings will induce 2-body proton decay (\ref{2bodyconstraint}). In the next section, we will discuss analogous constraints on $\left| \lambda^{\prime \prime}_{l m n} \lambda_{i j k }\right|$ for $l, m, n \neq 1, 1, 2$. Those bounds will be weaker than (\ref{4bodybound}) for reasons discussed therein. (\ref{4bodybound}) is valid for $\lambda_{1 3 2}$ and $\lambda_{2 3 1}$; $\lambda_{1 2 3}$ is more weakly bounded, as diagrams analogous to Figure \ref{Fig:4body} involving $\lambda_{1 2 3}$ would require Higgsino exchange or LR sfermion mixing in order to avoid a $\tau$ in the external legs.

Before proceeding, we discuss the relationship between the results in (\ref{4-bodydecay})-(\ref{4bodybound}) and previous work. Proton decay bounds on $\left|\lambda_{i j k} \lambda^{\prime \prime}\right|$ were briefly discussed in \cite{Carlson:1995ji}, which did not emphasize the antisymmetry condition $i \neq k, j \neq k$ and focused on the subdominant $p \rightarrow \nu \nu \ell^+$ decay mode (see Appendix \ref{3bodydecay} for further discussion). References \cite{Hoang:1997kf,Bhattacharyya:1998bx} first noted that $\lambda_{i j k }L_i L_j E^c_k, i \neq k, j \neq k$ induces $p \rightarrow K^+ \nu e^\pm \mu^\mp$ in the presence of B violating couplings. However, the dimensional analysis estimate of $\Gamma(p \rightarrow K^+ \nu e^\pm \mu^\mp)$ in \cite{Hoang:1997kf,Bhattacharyya:1998bx} did not account for hadronic matrix elements or phase space factors, resulting in a significantly stronger bound on $\left|\lambda^{\prime \prime}_{1 1 2} \lambda_{i j k}\right|$ than obtained here.

\subsection{Constraining Heavy Flavor $\lambda^{\prime \prime}$ Couplings}\label{heavyflavorconstraints}

In Section \ref{lightflavorconstraints}, we focused on finding bounds on $|\lambda^{\prime \prime}_{1 1 2}\lambda_{i j k }|$ for $i \neq k, j \neq k$. Similar bounds on $\lambda^{\prime \prime}_{1 1 3}, \lambda^{\prime \prime}_{1 2 3}, \lambda^{\prime \prime}_{2 1 2}$ and $\lambda^{\prime \prime}_{3 1 2}$ arise via tree-level diagrams involving flavor changing squark mass insertions; these bounds will be weaker than (\ref{4bodybound}) by a model dependent factor. Depending on these flavor violating parameters, different decay modes such as $p \rightarrow \nu_i \nu_j \ell^+_k$ or $n \rightarrow \nu_i \ell^+_j \ell^-_k$ may provide the dominant bound for these $\lambda^{\prime \prime}$ couplings, but this is a model dependent question which we will not address here.

\begin{figure}[t!]
\center{
\includegraphics[scale=0.9]{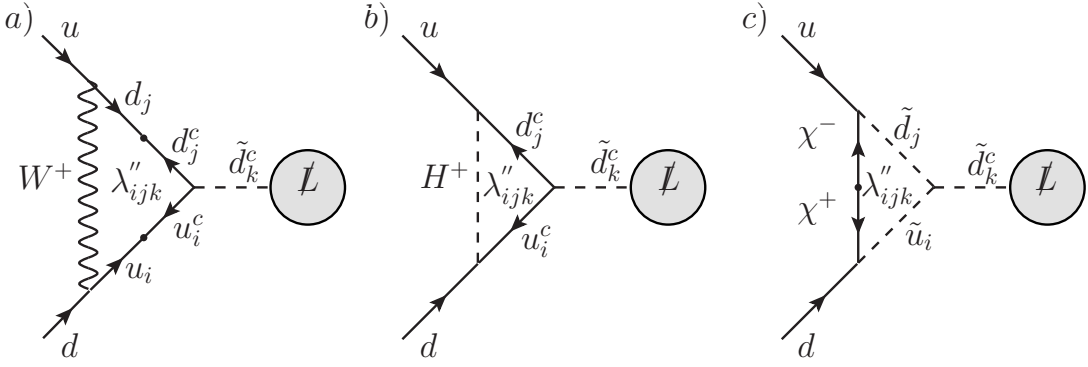}
\caption{\label{Fig:loops} Loop diagrams which generate proton-decay inducing effective operators for $\lambda^{\prime \prime}$ couplings with 2 heavy flavor indices \cite{Smirnov:1996bg}. The blob with $\slashed{L}$ denotes the L violating part of the diagram, akin to the right-hand side of Figure \ref{Fig:4body}.}}
\end{figure}

However, $\lambda^{\prime \prime}$ couplings with at least 2 heavy flavor ($c, b, t$) indices i.e. $\lambda^{\prime \prime}_{2 1 3}, \lambda^{\prime \prime}_{2 2 3}, \lambda^{\prime \prime}_{3 1 3}$ and $\lambda^{\prime \prime}_{3 3 2}$ will not contribute to nucleon decay at tree level, as tree-level diagrams analogous to Figure \ref{Fig:4body} will contain at least 2 heavy quark external legs. Instead, the relevant 6-fermion effective operators involving light quarks will be induced by flavor changing loop diagrams involving $W^{\pm}$, charged Higgs and chargino exchange, as depicted in Figure \ref{Fig:loops}.
 Diagrams of this sort were first discussed in \cite{Smirnov:1996bg}, though explicit formulae have yet to appear in the literature. 

Upon computing the coefficients of the 6-fermion operators from Figure \ref{Fig:loops}, we follow the procedure outlined in Section \ref{lightflavorconstraints} to compute nucleon decay rates. The $\lambda^{\prime \prime}_{2 3 1}$ and $\lambda^{\prime \prime}_{3 3 1}$ couplings generate effective operators which induce $n \rightarrow \nu_i \ell^-_j \ell^+_k$, while the $\lambda^{\prime \prime}_{2 3 2}$ and $\lambda^{\prime \prime}_{3 3 2}$ couplings generate effective operators which induce $p(n) \rightarrow K^+(K^0) \nu_i \ell^-_j \ell^+_k$. The resulting computation is slightly different from Section \ref{constraints} due to the operator structure of the diagrams in Figure \ref{Fig:loops}. The effective operators corresponding to the diagrams in Figure \ref{Fig:loops} are:
\begin{align}\label{6fermionloop}
\mathcal{L}_{6f} & \notag = \left(\lambda^{\prime \prime}_{2 3 1} L_{2 3 1} + \lambda^{\prime \prime}_{3 3 1} L_{3 3 1} \right)\lambda_{i j k} \left(\frac{2   \,g_{Y}^2}{3 M_{\tilde{B}} \tilde{m}_{d_R}^2}\right)\left(A_1  (u^\dagger d^\dagger) (d^c \ell^c_k) (\nu_i \ell_j) + A_2 (u^\dagger d^\dagger) (d^c \ell_j) (\nu_i \ell^c_k) \right)
\\ & +  \left(\lambda^{\prime \prime}_{2 3 2} L_{2 3 2} + \lambda^{\prime \prime}_{3 3 2} L_{3 3 2} \right)\lambda_{i j k} \left(\frac{2 g_{Y}^2}{3 M_{\tilde{B}} \tilde{m}_{s_R}^2}\right)\left(A_1 (u^\dagger d^\dagger) (s^c \ell^c_k) (\nu_i \ell_j) +  A_2 (u^\dagger d^\dagger) (s^c \ell_j) (\nu_i \ell^c_k)  \right),
\end{align} where again $A_1 =  1/(2 \tilde{m}_{\nu_i}^2) + 1/\tilde{m}_{{\ell_R}_k}^2$ and $A_2 = 1/(2 \tilde{m}_{\nu_i}^2)-1/(2 \tilde{m}_{{\ell_L}_j}^2)$. The $L_{i j k}$ are loop functions determined by summing the loop amplitudes and are derived in Appendix \ref{loopcalc}; the result is given in (\ref{loopfunction}). 

\begin{figure}[t!]
\center{
\includegraphics[width = 5.0 in]{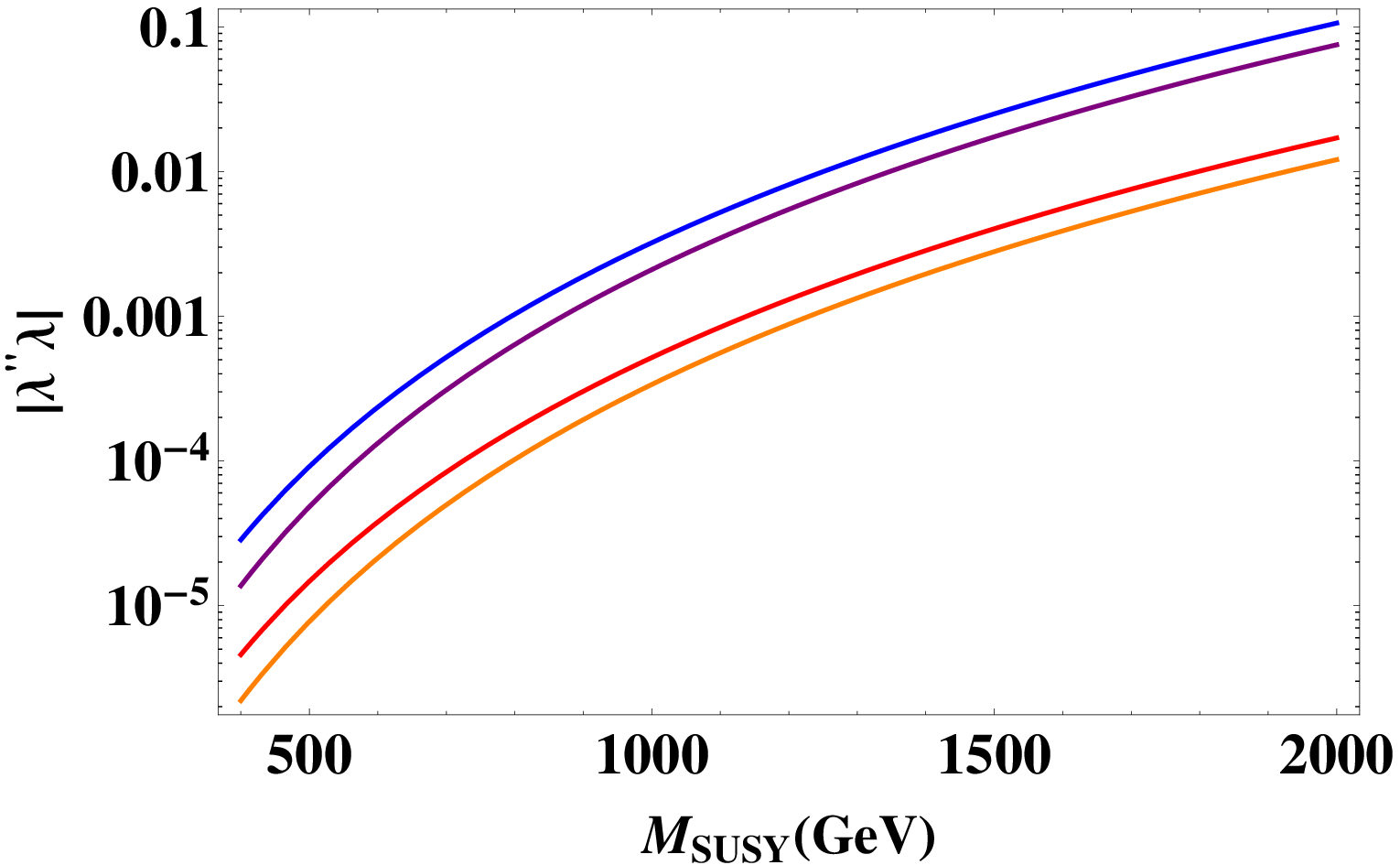}
 \caption{Bounds on $\left|\lambda^{\prime \prime}_{ l m n} \lambda_{i j k}\right|, i \neq k, j \neq k$ for $\lambda^{\prime \prime}$ couplings to $\ge 2$ heavy flavors, as a function of a common superpartner mass scale $M_{SUSY}$.  $\left|\lambda^{\prime \prime}_{2 3 2}\lambda_{ijk}\right|$ (blue) and $\left|\lambda^{\prime \prime}_{3 3 2}\lambda_{ijk}\right|$ (purple) are bounded by the 4-body nucleon decay modes $p(n) \rightarrow K^+(K^0) \nu_i \ell^-_j \ell^+_k$, while $\left|\lambda^{\prime \prime}_{2 3 1}\lambda_{ijk}\right|$ (red) and $\left|\lambda^{\prime \prime}_{3 3 1}\lambda_{ijk}\right|$ (orange) are bounded by the 3-body decay mode $n \rightarrow \nu_i \ell^-_j \ell^+_k$. For comparison, the corresponding nucleon decay bounds on $\left|\lambda^{\prime \prime}_{1 1 2} \lambda_{i j k}\right|$ are given in (\ref{4bodybound}).\label{Fig:bounds}}}
\end{figure}

The operators in (\ref{6fermionloop}) transform as elements in the ${\bf (3, \overline{3})}$ representation of $SU(3)_L \times SU(3)_R$, so in the degenerate sfermion ($A_2 = 0$) limit, the corresponding operators in the chiral Lagrangian are:
\begin{equation}
\mathcal{L} = C^\prime_1 \alpha \left({\ell^c_k}\, Tr\left[\mathcal{F}^{\prime} \xi {B^c} \xi\right]\right)\left({\nu_i} {\ell_j}\right)+ C^\prime_2\,\alpha \left({\ell^c_k}\, Tr\left[\tilde{\mathcal{F}}^{\prime \prime} \xi {B^c} \xi\right]\right)\left({\nu_i} {\ell_j}\right).
\end{equation} Here $\alpha$ is defined as $\left<0| (u_L d_L) u_R | p \right> \approx \alpha P_R u$ in Dirac notation; lattice calculations give $\alpha \approx \beta \approx 0.0118 \, (\mathrm{GeV})^3$ \cite{Aoki:2006ib}. Matching the operators in (\ref{6fermionloop}) to the chiral Lagrangian gives :\begin{equation} C^\prime_1 = \frac{A g_{Y}^2 \lambda_{i j k}}{M_{\tilde{B}} {\tilde{m}}^4} \left(\lambda^{\prime \prime}_{2 3 1} L_{2 3 1} + \lambda^{\prime \prime}_{3 3 1} L_{3 3 1} \right), \,\,\,\,C^\prime_2 = \frac{A g_{Y}^2 \lambda_{i j k}}{M_{\tilde{B}} {\tilde{m}}^4}\left(\lambda^{\prime \prime}_{2 3 2} L_{2 3 2} + \lambda^{\prime \prime}_{3 3 2} L_{3 3 2} \right)\end{equation} where $\tilde{m}$ is again the degenerate sfermion mass. The operator with coefficient $C^\prime_2$ induces the decay modes $N \rightarrow K \nu_i \ell^-_j \ell^+_k$ whose rate (neglecting lepton masses) is given by equation (\ref{4-bodydecay}), with \begin{equation}C_N = \frac{C^\prime_2\alpha}{f_\pi} \left(1 + \frac{\left(D + 3 F\right) M_N}{3 M_\Lambda}\right) \cite{Aoki:1999tw}. \end{equation} The operator with coefficient $C^\prime_1$ induces the 3-body neutron decay mode $n \rightarrow \nu_i \ell^-_j \ell^+_k$, whose rate is given by:
\begin{equation}
\Gamma(n \rightarrow \nu_i \ell^-_j \ell^+_k)= \frac{\alpha^2 \left|C^\prime_1\right|^2}{6144 \pi^3} {M_n}^5
\end{equation}
The bound on the partial width for $n \rightarrow \nu e^+ \mu^-$ from IMB-3 is similar to the bound on $N \rightarrow \,\mu^+ \, \mathrm{inclusive}$, $\tau(n \rightarrow \nu e^+ \mu^-) \lesssim 10^{32}$ years \cite{McGrew:1999nd}. Comparing the computed decay rates with these constraints, we obtain bounds on $\left|\lambda^{\prime \prime}_{ l m n} \lambda_{i j k}\right|, i \neq k, j \neq k$, for $(l, m, n) = (2,3,1),\, (3, 3, 1), \,(2, 3, 2)$ and $(3, 3, 2)$, which are plotted in Figure \ref{Fig:bounds} as a function of a common superpartner mass scale $M_{SUSY}$. These bounds are significantly weaker than (\ref{4bodybound}), as the diagrams in Figure \ref{Fig:loops} are suppressed by quark Yukawa couplings and off-diagonal $V_{CKM}$ elements, along with the usual $1/(16\pi^2)$ loop factor. We remind the reader that Figure \ref{Fig:bounds} only applies to $\lambda_{1 3 2}$ and $\lambda_{2 3 1}$; bounds on $\lambda_{1 2 3}$ are weaker for reasons mentioned above.

We close this section by remarking that discovery of the nucleon decay modes $N \rightarrow K e^\pm \mu^\mp \nu$ and $n \rightarrow e^\pm \mu^\mp \nu$ without a similar discovery in same flavor lepton modes e.g. $N \rightarrow K e^+ e^- \nu$, $n \rightarrow e^+ e^- \nu$ would provide strong evidence for BLRPV. Note that bounds quoted above on $\tau(N \rightarrow \mu^+ + \, \mathrm{anything})$\cite{Cherry:1981uq} and $\tau(n \rightarrow \nu e^+ \mu^-)$\cite{McGrew:1999nd} are more than a decade old. We urge experimentalists to continue searching for these decay modes, as the nucleon decay signatures of BLRPV are significantly more robust and model independent than the collider signatures discussed below (see Section \ref{collider}).

\section{The Flavor Symmetry of BLRPV}\label{flavorsym}

Having seen that a hierarchy between $\lambda_{i j k} L_i L_j E^c_K, i \neq k, j \neq k$ and all other LRPV couplings avoids the 2-body proton decay bounds in (\ref{2bodyconstraint}), we now consider the question of whether such a hierarchy can be obtained naturally. This might seem difficult, because once lepton number is violated, there are usually no remaining symmetries which can distinguish the lepton chiral multiplets $L_i$ from the Higgs multiplet $H_d$. As a result, the presence of LRPV couplings  typically induces wavefunction renormalization mixing of $L$ and $H_d$ \cite{Hall:1983id,deCarlos:1996du}, which, for example, radiatively generates a $\lambda^{\prime} Q L D^c$ coupling proportional to the down-type Yukawa couplings. 

However, even if lepton number is violated, there can still be flavor symmetries acting in the lepton sector which forbid certain LRPV operators. Neglecting neutrino masses, if $\lambda_{i j k} L_i L_j E^c_k$ for $i \neq k, j \neq k$ are the only non-vanishing LRPV couplings in the basis where $Y_E$ is diagonal, the theory enjoys a global $Z_2^e \times Z_2^\mu \times Z_2^\tau$ flavor symmetry. Under $Z_2^\tau$, both $L_\tau$ and $E^c_\tau$ are even while all other lepton fields are odd; $Z_2^\mu$ and $Z_2^e$ are similarly defined. Thus absent neutrino masses, the vanishing of all other $L$ violating couplings is protected by this global symmetry. $Z_2^e \times Z_2^\mu \times Z_2^\tau$ also forbids all 4-fermion operators of the form $q q q \ell$, explaining the absence of 2-body proton decay from an effective operator point of view.

\begin{figure}[t!]
\center
\includegraphics{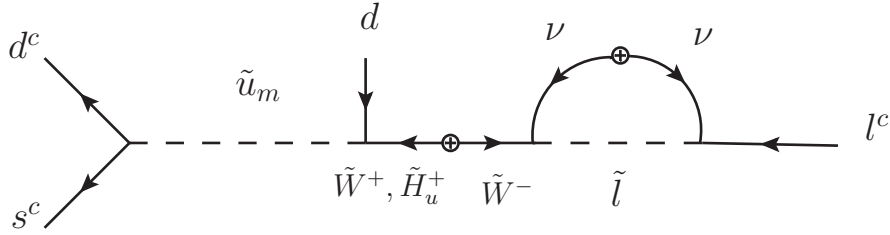}
\caption{Loop diagram which generates an effective $(d^c s^c)^\dagger (d \ell^c)$ operator in the presence of non-vanishing neutrino masses and mixing.\label{Fig:Section3Fig}}
\end{figure}

In models with a realistic neutrino sector, $Z_2^e \times Z_2^\mu \times Z_2^\tau$ will be broken by neutrino masses and mixing angles \cite{Leurer:1992wg}. Thus we expect the presence of non-vanishing neutrino masses to induce effective 4-fermion $q q q \ell$ operators\footnote{We thank Aaron Pierce for emphasizing this point.} whose coefficients are proportional to $m_\nu$. Such operators are indeed generated, by loop diagrams which induce lepton-gaugino/higgsino mixing. The dominant diagram of this sort is depicted in Figure \ref{Fig:Section3Fig}, resulting in the following effective operator:\begin{equation}\label{neutrinomassoperator}
\mathcal{L} \supset \left(\frac{\lambda^\prime_{eff} \lambda^{\prime \prime}_{m 1 2}}{{M_{SUSY}}^2}\right) \left( d^c s^c\right)^\dagger \left(d \, \ell^c\right), \hspace{4mm} \lambda^\prime_{eff} \sim \frac{y_{u_m}V_{m 1}\lambda_{i j k } m_{\nu}}{16 \pi^2 M_{SUSY}}, \end{equation} which induces the decay mode $n \rightarrow K^\pm \ell^\mp$. $M_{SUSY}$ is taken to be the common superpartner mass scale. We have verified that other similar diagrams give $q q q \ell$ operators whose coefficients are suppressed with respect to (\ref{neutrinomassoperator}) by lepton mass insertions, LR squark mixing, and/or powers of spacetime derivatives. Given the bound on $\left|\lambda^{\prime \prime} \lambda^\prime\right|$ in (\ref{2bodyconstraint}), the resulting bound on $\left|\lambda^{\prime \prime}_{1 1 2} \lambda_{i j k }\right|$ from (\ref{neutrinomassoperator}) is weaker than (\ref{4bodybound}) for $M_{SUSY} \lesssim 100$ TeV for $m_\nu \sim 0.1$ eV.

Thus if neutrino masses are the only source of  $Z_2^e \times Z_2^\mu \times Z_2^\tau$  breaking, bounds from 2-body proton decay are mild. However, the same dynamics which breaks $Z_2^e \times Z_2^\mu \times Z_2^\tau$ and generates neutrino masses might also regenerate other dangerous LRPV operators. This is a model dependent issue, which depends on the UV dynamics responsible for neutrino mass generation. As an illustrative example, we analyze a particular right-handed neutrino model, where $Z_2^e \times Z_2^\mu \times Z_2^\tau$ is embedded within a spurious $SU(3)_{\ell} \times SU(3)_N$ flavor symmetry that is broken by lepton and neutrino Yukawa couplings (see \cite{Csaki:2011ge} for a similar analysis). Bounds from 2-body proton decay will constrain the right-handed neutrino sector, but can still allow for right-handed neutrinos above the TeV scale, assuming non-holomorphic contributions to the superpotential are sufficiently suppressed. The details of this model and the resulting analysis are presented in Appendix \ref{flavor}.

\section{Collider Phenomenology of BLRPV}\label{collider}

In this section, we discuss how the collider phenomenology of BLRPV can be qualitatively different from that of canonical BRPV or LRPV. In BLRPV, sparticles can have both BRPV and LRPV decay modes, in addition to the usual R-parity conserving decay modes. The resulting collider phenomenology then depends on  branching ratios sparticles to BRPV and LRPV final states. For clarity, we separate the novel phenomenology of BLRPV into two distinct scenarios:\begin{itemize}
\item \emph{Scenario A}: $\tilde{X} \rightarrow \, \mathrm{BRPV},\, \tilde{X} \rightarrow \, \mathrm{LRPV}$. A given sparticle $\tilde{X}$ decays to both BRPV and LRPV final states with comparable branching fractions.
\item \emph{Scenario B}: $\tilde{X}_1 \rightarrow \, \mathrm{BRPV}$, $\tilde{X}_2 \rightarrow \, \mathrm{LRPV}$. A given sparticle $\tilde{X}_1$ decays predominantly via LRPV, while a different sparticle $\tilde{X}_2$ decays predominantly via BRPV. \end{itemize} The dichotomoy between Scenarios A and B is somewhat artificial, as the branching ratios to LRPV and BRPV final states for each sparticle can take on a continuum of values in an arbitrary model. Nevertheless we discuss each case separately, and highlight how each scenario can give rise to novel collider phenomenology from sparticle production. We then discuss implications of existing collider constraints on these BLRPV scenarios. A similar study of B and L violating collider signatures was performed in \cite{Durieux:2012gj} from an effective field theory point of view.

\subsection*{Scenario A: $\tilde{X} \rightarrow \, \mathrm{ BRPV},\, \tilde{X} \rightarrow \, \mathrm{LRPV}$}

In this scenario, pair production of $\tilde{X}$ can lead to novel final states if one $\tilde{X}$ decays via LRPV and the other $\tilde{X}$ decays via BRPV. This can be realized if for instance $\tilde{X}$ is a neutralino/chargino\footnote{If there is a hierarchy between the LRPV and BRPV couplings, this scenario can also occur for squark, slepton and gluino LSP's.} (co)LSP, and the antisymmetric LRPV and BRPV couplings are similar in magnitude. 

Suppose $\tilde{X}$ is a Wino-like LSP, in which case there is a nearly mass degenerate chargino NLSP $\chi^+$. For  approximately degenerate sfermions and sufficiently large $\gtrsim 10^{-5}$ BRPV and LRPV couplings, $\chi^+$ will have a sizeable branching fraction to both $\chi^+ \rightarrow q q q$ and $\chi^+ \rightarrow e \mu \tau$ if $\lambda^{\prime \prime} \sim \lambda$. This leads to new final states from colored sparticle production:\begin{equation}\label{finalstates}
\tilde{q} \tilde{q} \rightarrow 2 q {\chi}^+ {\chi}^- \rightarrow 5 q \, e\, \mu \,\tau,\hspace{5mm} \tilde{g} \tilde{g} \rightarrow 4 q\, {\chi}^- {\chi}^+ \rightarrow 7 q \, e \,\mu \,\tau,\hspace{5mm} \tilde{g} \tilde{q} \rightarrow 3 q\, {\chi}^+ {\chi}^- \rightarrow 6 q \, e \,\mu \,\tau
\end{equation} Such final states with large jet multiplicity, 3 hard leptons of different flavors and no MET can differentiate BLRPV from both the R-parity conserving and BRPV/LRPV MSSM. This example illustrates the general point that sparticle pair production in Scenario A leads to final states with large jet multiplicity and $\left|\Delta L \right| = 1$, where the leptons will be of different flavors due to the antisymmetric flavor structure in the LRPV couplings.

Note that the above signatures (\ref{finalstates}) can also be mimicked by pure LRPV. For instance, squark pair production and decay via $L L E^c$ can result in final states such as $\tilde{q} \tilde{q} \rightarrow q q \chi^0 \chi^+ \rightarrow q q  e \mu \tau + \nu\mathrm{'s}$ with additional jets from QCD radiation, giving final states similar to (\ref{finalstates}). It is non-trivial to experimentally distinguish these final states from the BLRPV signatures in (\ref{finalstates}). However, pure LRPV final states with an $e, \mu$ and $\tau$ must be accompanied by an additional neutrino(s). This is because sparticle pair production and decay via pure LRPV will result in final states which violate L in even units. Thus depending on the particular spectrum in question, missing transverse energy from neutrinos can potentially be used to distinguish pure LRPV from the BLRPV signatures in (\ref{finalstates}).

\subsection*{Scenario B: $\tilde{X}_1 \rightarrow \, \mathrm{LRPV}$, $\tilde{X}_2 \rightarrow \, \mathrm{BRPV}$} 

\begin{figure}[t!]
\center{
\includegraphics[scale=0.75]{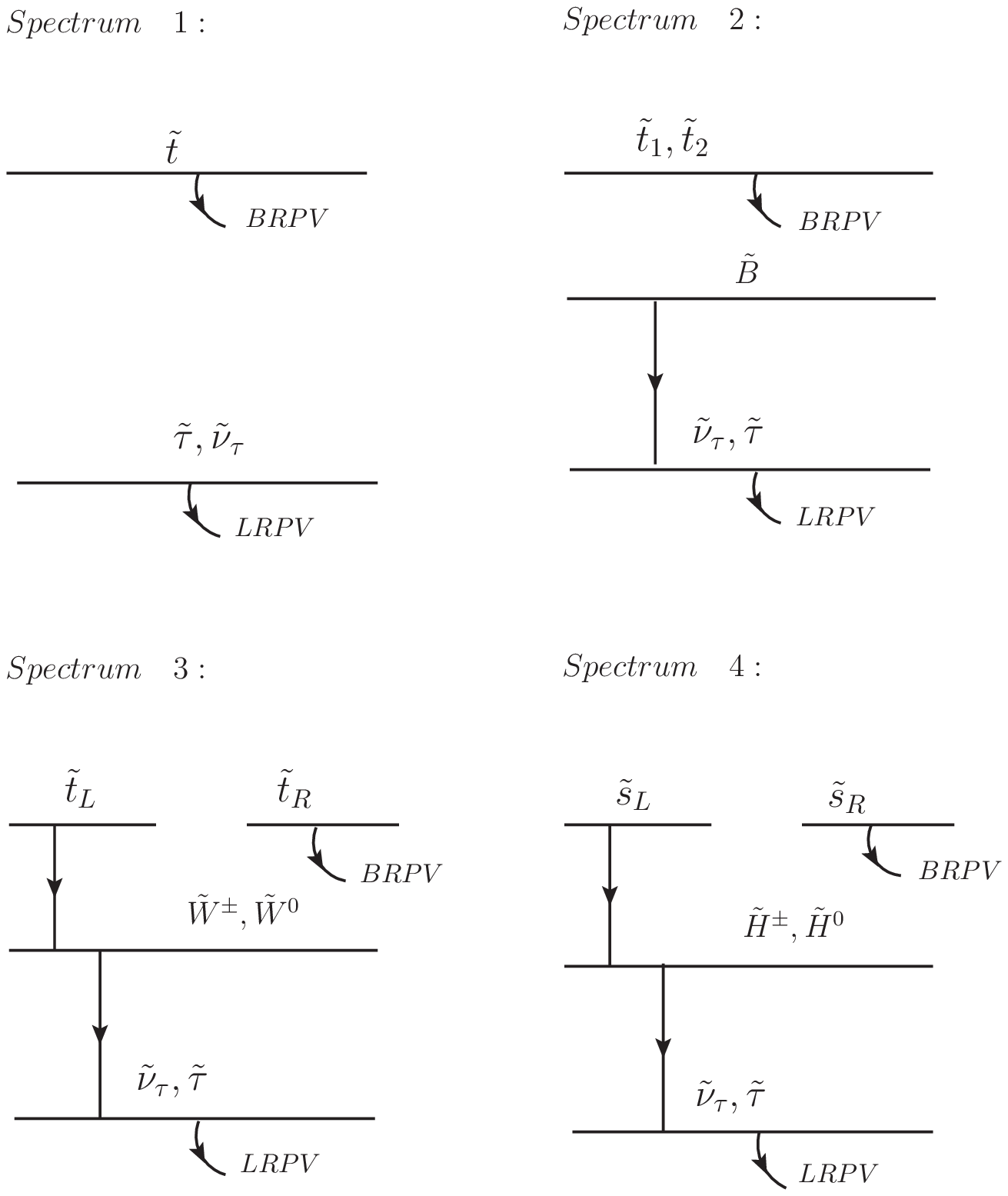}
 \caption{Simplified models with $\tilde{\nu}_{\tau}$ LSP discussed in the text. $\left|\lambda^{\prime \prime}_{3 3 2} \right| = 0.1$ is assumed to be the only nonvanishing BRPV coupling.\label{spectra}}}
\end{figure}

There are numerous qualitatively distinct possibilities in Scenario B, depending on the mass spectrum and the magnitudes of the LRPV and BRPV couplings. In this work we will focus on four particular examples involving the simplified spectra depicted in Figure \ref{spectra}. These examples are meant to illustrate generic features of models which fall into this scenario; we save a discussion of the more general case for future work. The four examples in Figure \ref{spectra} all feature a squark decaying via BRPV, a $\tilde{\tau}_L$ NLSP, and a $\tilde{\nu}_\tau$ LSP which decays via the antisymmetric LRPV couplings to $\tilde{\nu}_\tau \rightarrow \mu e$. Our choice of a $\tilde{\nu}_\tau$ LSP is motivated by the fact that displaced decays to a $\mu e$ pair can mitigate constraints from existing RPV searches\cite{Batell:2013bka}. We return to this point in Section \ref{displaced}.

In discussing the examples of Figure \ref{spectra}, we assume that $\left|\lambda^{\prime \prime}_{3 3 2}\right| = 0.1$, and all other BRPV couplings are negligible. This ensures that 3-body decay modes such as as $\tilde{q} \rightarrow q^\prime \, \tau\, \tilde{\nu}_\tau, \, \tilde{q} \rightarrow q \, \nu_\tau \tilde{\nu}_\tau$ can be subdominant to squark BRPV decay modes, regardless of the virtual neutralino/chargino masses. The dominant constraint\footnote{Preserving a primordial matter asymmetry imposes a much stronger constraint on $\lambda^{\prime \prime}_{3 3 2}$\cite{Campbell:1990fa}. However these constraints can be avoided if e.g. baryogenesis takes place below the electroweak phase transition.} on $\lambda^{\prime \prime}_{3 3 2 }$ is $\left|\lambda^{\prime \prime}_{3 3 2} \right| \lesssim 1$, which arises from requiring perturbativity up to the GUT scale \cite{Brahmachari:1994wd,Allanach:1999ic}; our fiducial value is well below this bound. We now discuss each example case by case:\begin{itemize}

\item \textbf{Spectrum 1: Light Stop}. This example is fairly self-explanatory. Depending on the magnitude of stop mixing, either one or both stop mass eigenstates decay via BRPV,

\item \textbf{Spectrum 2: Light Bino}. In this example, we assume large stop mixing i.e. $\sin \theta_{\tilde{t}} \sim 0.5$ and $M_{\tilde{t}} - M_{\tilde{B}} \lesssim m_t$. The R-parity conserving mode $\tilde{t} \rightarrow t \, \tilde{B}$ is kinematically forbidden, so the stops decay predominantly via BRPV to $\tilde{t} \rightarrow \overline{b} \, \overline{s}$.

\item \textbf{Spectrum 3: Light Wino}. In this example, we assume vanishing LR stop mixing. The dominant contributions to the R-parity conserving decay modes $\tilde{t}_R \rightarrow t \tilde{W}^0, b \tilde{W}^+$ arise from Higgsino-Wino mixing. Taking the $\mu \gg M_Z, M_2$  limit, the relevant mixing angles are $\sqrt{2} M_W \left(\mu c_\beta + M_2 s_\beta \right)/\mu^2$ and $M_Z \cos \theta_W/\sqrt{2} \mu$ in the chargino and neutralino sectors. Taking $\mu \gtrsim 1$ TeV, $M_2 \sim 200$ GeV and $\tan \beta \sim 10$, the BRPV decay mode $\tilde{t}_R \rightarrow \overline{b} \overline{s}$ dominates over the R-parity conserving decay modes for $\tilde{t}_R$.

\item \textbf{Spectrum 4: Light Higgsino}. For this example we assume moderate $\tan\beta$ such that $Y_s \lesssim 10^{-2}$. The dominant R-parity conserving decay mode for $\tilde{s}_R$ is $\tilde{s}_R \rightarrow s\,\tilde{H}^0$ with a contribution from Higgsino-Bino mixing with mixing angle $M_Z \sin \theta_W/\sqrt{2} M_1$ in the $M_1 \gg M_Z, \mu$ limit. Thus for $M_1 \gtrsim 1$ TeV, the BRPV mode $\tilde{s}_R \rightarrow \overline{t} \, \overline{b}$ dominates. The BRPV decay mode for $\tilde{s}_L$ is suppressed by LR mixing, so $\tilde{s}_L$ will decay predominantly via R-parity conserving channels.

\end{itemize}

In these examples, sparticle pair production and associated $\tilde{g} \tilde{q}$ production\footnote{If the $\tilde{g}$ is heavier (lighter) than $\tilde{q}$, the gluino (squark) will decay via the QCD coupling to $\tilde{g} \rightarrow q \tilde{q}$ ($\tilde{q} \rightarrow q \tilde{g}$), resulting in the same final state as pair production with an additional jet.} yield final states identical to that of pure BRPV or pure LRPV, depending on the sparticle produced. Thus to distinguish Scenario B from standard RPV, an experimental discovery in at least two different channels would be required. For instance, if any of the Spectrums 1-3 with vanishing stop mixing is realized in nature, the discovery of a RH stop decaying via BRPV along with a LH stop decaying via LRPV would give conclusive evidence for BLRPV.

\subsection*{Collider Constraints on BLRPV}\label{colliderconstraints}

We now review existing LHC searches for standard RPV scenarios, which are also sensitive to BLRPV scenarios. For BRPV, ATLAS and CMS have recently released searches for purely hadronic final states \cite{ATLAS-CONF-2013-091,Chatrchyan:2013gia} which place gluino mass bounds of $\sim 800 \,(950)$  GeV, assuming decoupled (light) neutral/charged-inos and decoupled squarks. Squarks decaying to 2 jets via BRPV are still weakly constrained, though lighter squarks can push the gluino mass limit up to $\gtrsim 1.5$ TeV \cite{Graham:2014vya}. For LRPV, CMS and ATLAS searches in multilepton final states \cite{Chatrchyan:2013xsw,CMS-PAS-SUS-12-027,CMS-PAS-SUS-13-010,Aad:2014iza} have placed bounds of $\gtrsim 1$ TeV for stop masses, $\gtrsim 1.5$ TeV for gluino and other squarks masses, and $\gtrsim 750$ GeV for charginos, assuming LSP decays give prompt, isolated leptons. The quoted bounds are for simplified models; more complicated spectra which produce additional hard objects through cascade decays are significantly more constrained by these searches \cite{Evans:2012bf,Asano:2012gj,Evans:2013jna}.

It is straightforward to interpret these searches for the BLRPV scenarios discussed above. For Scenario A, $\tilde{X}$ pair production will yield events where both $\tilde{X}$ particles decay via the same BRPV or LRPV decay mode. Thus bounds on $\tilde{X}$ production will be similar to bounds on standard BRPV/LRPV scenarios, albeit slightly weaker due to branching ratio suppression of the effective cross section. For Scenario B with $\tilde{X}_1 \rightarrow$ LRPV and $\tilde{X}_2 \rightarrow$ BRPV, the above searches are also straightforward to apply, as pair production of either $\tilde{X}_1$ or $\tilde{X}_2$ gives signatures identical to that of standard RPV.

Though we have focused on how BLRPV can give unique phenomenology, it is also possible for BLRPV to give collider signals identical to that of standard RPV scenarios. In this case, one would have to rely on the nucleon decay phenomenology discussed in Section \ref{constraints} to differentiate BLRPV from standard RPV scenarios.

\section{Displaced $\tilde{\nu}_{\tau} \rightarrow \mu e$ Decays in BLRPV}\label{displaced}

We now discuss the phenomenology of displaced $\tilde{\nu}_\tau \rightarrow \mu e $ decays, where $c\tau({\tilde{\nu}_\tau}\rightarrow \mu e  ) \sim \mathcal{O}(\mathrm{mm}) - \mathcal{O}(\mathrm{m})$. This might seem unrelated to our discussion of BLRPV thus far, as the $\tilde{\nu}_\tau \rightarrow \mu e$ decay mode only involves LRPV couplings. However, the $Z_2^e \times Z_2^\mu \times Z_2^\tau$ flavor symmetry of BLRPV gives a natural setting for displaced $\tilde{\nu}_\tau \rightarrow \mu e$ decays if the LSP is a $\tau$-sneutrino, due to the antisymmetric flavor structure in the $L L E^c$ couplings. Thus although much of the following discussion is not unique to BLRPV, models with displaced $\tilde{\nu}_\tau \rightarrow \mu e$ decays can arise naturally in BLRPV, compared to other RPV scenarios that do not motivate such a flavor structure in the LRPV couplings.

In general LRPV scenarios where the LSP has a has a macroscopic decay length i.e. $c \tau \sim \mathcal{O}(\mathrm{mm}) - \mathcal{O}(\mathrm{m})$, leptons resulting from LSP decay will fail impact parameter cuts imposed by the LRPV searches \cite{Chatrchyan:2013xsw,CMS-PAS-SUS-12-027,CMS-PAS-SUS-13-010,Aad:2014iza} referenced in Section \ref{collider}. Consequently, the strict kinematic bounds from these searches will no longer apply if the LSP decay is displaced \cite{Graham:2012th}. Displaced $\tilde{\nu}_\tau \rightarrow \mu e$ decays are particularly noteworthy, as they also evade dedicated LHC searches for displaced final states  \cite{Chatrchyan:2012jna,ATLAS-CONF-2013-092,CMS-PAS-EXO-12-038}\footnote{CMS has recently released a search for events with a \emph{single} displaced $\mu$ and $e$ \cite{CMS-PAS-B2G-12-024}, which has limited sensitivity to events with two sneutrinos undergoing displaced decays to $\mu e$.}, and for $c\tau({\tilde{\nu}_\tau}\rightarrow \mu e ) \gtrsim 10$ cm can even evade bounds from LEP searches \cite{Batell:2013bka}. Thus the spectra depicted in Figure \ref{spectra} are relatively unconstrained by collider searches if the $\tau$-sneutrino decay is displaced, as collider constraints on squarks decaying to 2 jets are rather weak ($m_{\tilde{t}} \gtrsim 200$ GeV \cite{Bai:2013xla}). Squarks which cascade decay to $\tilde{\nu}_{\tau}$ via an intermediate neutralino/chargino can be more strongly constrained if $\chi^+ \rightarrow \chi^0$ decays yield additional hard objects in the final state.

\begin{figure}
\centering
\includegraphics[scale=0.57]{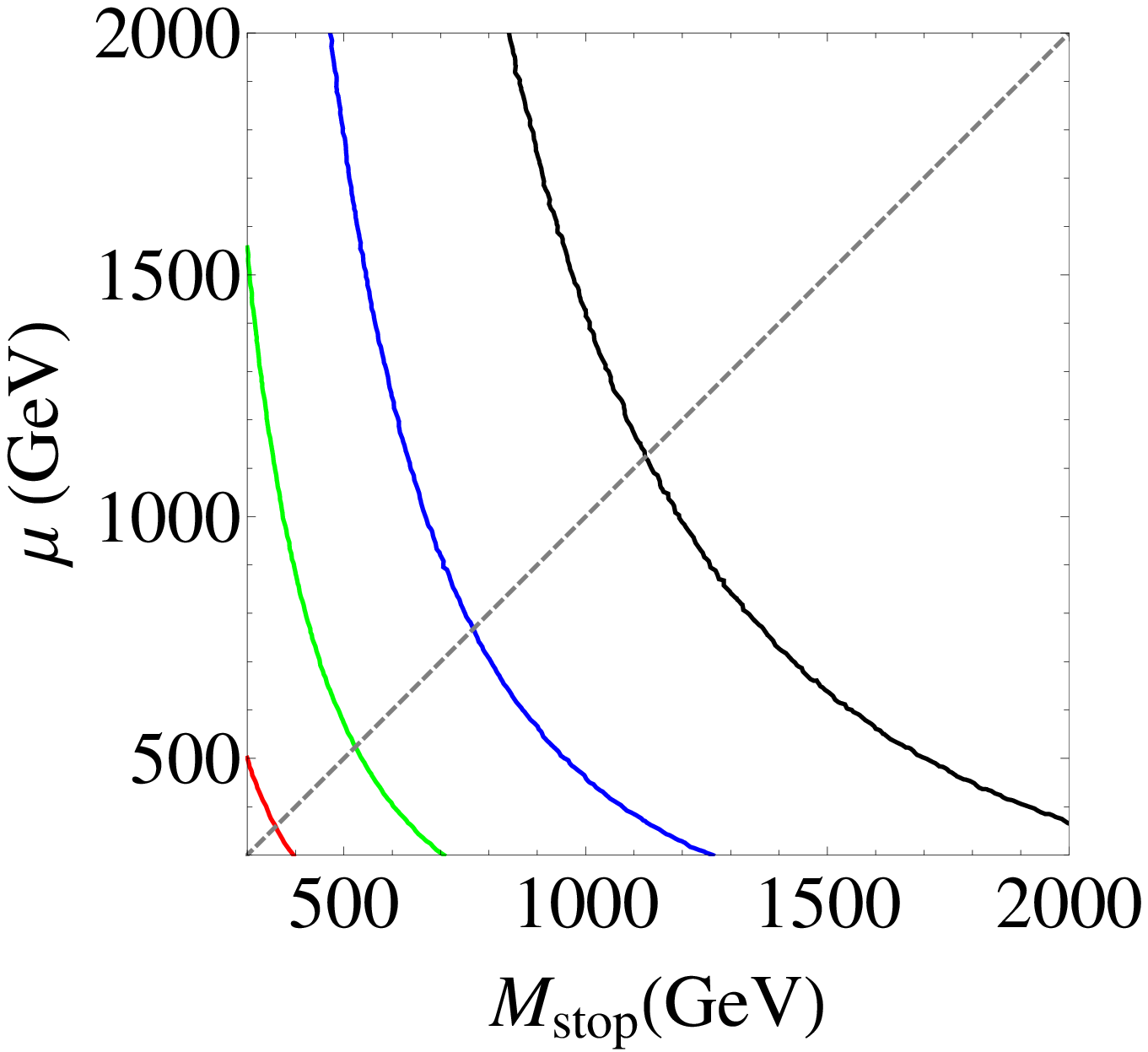}
\includegraphics[scale=0.57]{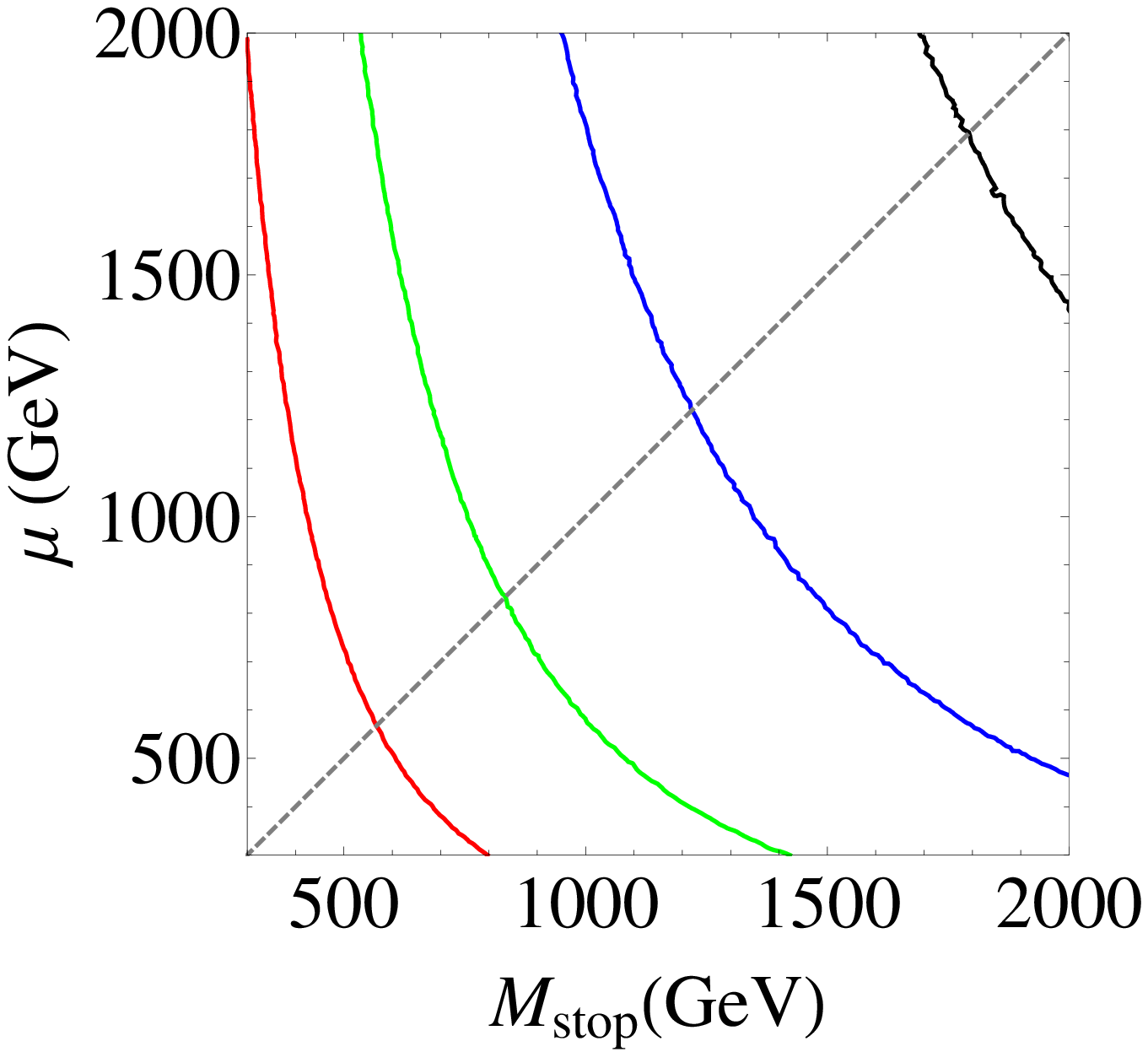}
\caption{Contours of constant $c \tau(\tilde{\nu}_\tau \rightarrow \tau \overline{b}\, \overline{b} \,\overline{s})$ for $m_{\tilde{\nu}_\tau} = 100$ GeV, $\left|\lambda^{\prime \prime}_{3 3 2}\right| = 0.1$, $M_2 = 300$ GeV. The left (right) plot corresponds to $\tan \beta =$ 2 (10). (Red, Green, Blue, Black) curves correspond to $c \tau(\tilde{\nu}_\tau \rightarrow \tau \overline{b}\, \overline{b} \,\overline{s}) = (0.1, 1, 10, 100)$ meters; the dashed line represents $M_{stop} = \mu$.\label{sneutrinoplot}}
\end{figure}

For the BRPV coupling assumed in Figure \ref{spectra} ($\left|\lambda^{\prime \prime}_{3 3 2}\right| = 0.1$), the $\tau$-sneutrino has the BRPV decay modes $\tilde{\nu}_\tau \rightarrow \tau \overline{b} \,\overline{b} \,\overline{s}$ and  $\tilde{\nu}_\tau \rightarrow \nu\, t \,b \,s$. These decay modes can compete with $\tilde{\nu}_\tau \rightarrow \mu e$, particularly if the decay length for $\tilde{\nu}_\tau \rightarrow \mu e$ is macroscopic. Focusing on the spectra in Figure \ref{spectra}, we compute the $\tilde{\nu}_\tau \rightarrow \tau \overline{b}\, \overline{b}\, \overline{s}$ decay width in Appendix \ref{4bodysneutrinodecay}, which is the dominant BRPV decay mode if $m_{\tilde{\nu}_\tau} \lesssim m_t$. The dominant diagrams for this decay mode (shown in Figure \ref{sneutrino} of Appendix \ref{4bodysneutrinodecay}) involve virtual higgsino and stop exchange; thus $\Gamma(\tilde{\nu}_\tau \rightarrow \tau \overline{b}\, \overline{b}\, \overline{s})$ is rather sensitive to $\mu$ and $m_{\tilde{t}}$. The BRPV decay width for $\tilde{\nu}_{\tau}$ in the $\mu$-$m_{\tilde{t}}$ plane is depicted in Figure \ref{sneutrinoplot}, which takes $\left|\lambda^{\prime \prime}_{3 3 2}\right| = 0.1$, $M_2 = 300$ GeV and $m_{\tilde{\nu}} = 100$ GeV. We assume vanishing stop mixing, and take the degenerate stop limit $m_{\tilde{t}_1} = m_{\tilde{t}_2} = m_{\tilde{t}}$. We see that there are large regions of parameter space with $m_{\tilde{t}}, \mu \lesssim 2$ TeV and $c\tau(\tilde{\nu}_\tau \rightarrow \tau \overline{b} \,\overline{b} \,\overline{s}) \gtrsim 1$ meter, particularly for small $\tan \beta$ (we have defined $c\tau(\tilde{\nu}_\tau \rightarrow \tau \overline{b}\, \overline{b}\, \overline{s}) \equiv c/\Gamma(\tilde{\nu}_\tau \rightarrow \tau \overline{b}\, \overline{b}\, \overline{s})$). Thus if $c\tau(\tilde{\nu}_\tau \rightarrow \mu e) \sim \mathcal{O}(\mathrm{cm})$, the branching ratio to $\tilde{\nu}_\tau \rightarrow \tau \overline{b}\, \overline{b}\, \overline{s}$ can be subdominant, even for the large BRPV couplings ($\left|\lambda^{\prime \prime}_{3 3 2}\right| = 0.1$) assumed in Figure \ref{spectra}. To the best of our knowledge the calculation of Appendix \ref{4bodysneutrinodecay}, which is also relevant for studying pure BRPV with a slepton LSP, has not appeared elsewhere in the literature. 

\section{Conclusion}\label{conclusion}

We have established here a class of RPV models which violate B and L simultaneously (BLRPV), without inducing unacceptable nucleon decay. BLRPV requires an approximate $Z_2^e \times Z_2^\mu \times Z_2^\tau$ flavor symmetry in the lepton sector, which forbids 4-fermion effective operators leading to 2-body nucleon decay. This symmetry also forbids all LRPV operators aside from $\lambda_{i j k } L_i L_j E^c_k$ for $i \neq k, j \neq k$, significantly reducing the number of free parameters usually associated with the LRPV superpotential. Nucleons are predicted to decay through the decay modes $N \rightarrow K \nu e^\pm \mu^\mp$ and $n \rightarrow \mu^\pm e^\mp \nu$. A discovery of nucleon decay in these modes, without discoveries in similar modes with same flavor leptons, would give a smoking gun signature for BLRPV.

Current nucleon lifetimes bounds on BRPV and  $\lambda_{i j k } L_i L_j E^c_k$, $i \neq k, j \neq k$ couplings are rather weak, allowing both to be relevant for collider phenomenology. Novel phenomenology arises in BLRPV because sparticles can decay via both LRPV and BRPV couplings. Exotic final states can arise from sparticle pair production, if one sparticle decays through BRPV while the other through LRPV. These final states are characterized by large jet multiplicity, three hard leptons of different flavor and no missing energy, e.g. $\tilde{q} \tilde{q} \rightarrow 2 q \chi^+ \chi^- \rightarrow 5 q e \mu \tau$. Alternatively, different sparticles could decay predominantly via either BRPV or LRPV, allowing both pure BRPV and pure LRPV signals to manifest within the same spectrum. 

Due to the flavor structure in $L L E^c$ couplings enforced by $Z_2^e \times Z_2^\mu \times Z_2^\tau$, BLRPV provides a natural framework for displaced $\tilde{\nu}_\tau \rightarrow \mu e $ decays to occur, provided $\tilde{\nu}_\tau$ is the LSP. This decay mode allows sleptons and charginos to evade constraints from both LEP and LHC searches for LRPV \cite{Batell:2013bka}. We have demonstrated that even if BRPV couplings are large (e.g. $\left|\lambda^{\prime \prime}_{3 3 2}\right| \sim 0.1$), a $\tau$-sneutrino LSP can still decay predominantly via $\tilde{\nu}_{\tau} \rightarrow \mu e $. This allows for spectra in which e.g. squarks decay to jets via the $U D^c D^c$ operator while charginos/sleptons decay through $\tilde{\nu}_\tau \rightarrow \mu e$ via $L L E^c$. Such spectra are weakly constrained by existing collider searches compared to other R-parity violating and R-parity conserving SUSY scenarios.

Although we have focused on RPV supersymmetry in this work, our analysis here (particularly in Section \ref{constraints}) illustrates the general constraint that 6-fermion operators of the form $q q q \ell \ell \ell/\tilde{\Lambda}^5$ must satisfy $\tilde{\Lambda} \gtrsim 100\,(10)$ TeV for couplings to $u, d, s$ ($c, b, t$) quarks. Therefore, interactions of TeV scale particles can violate B and L without violating nucleon decay bounds, provided some structure is in place to suppress 4-fermion operators of the form $q q q \ell/\tilde{\Lambda}^2$; a similar observation was made by Weinberg in \cite{Weinberg:1980bf}. The explicit computation of nucleon decay rates from these effective operators has not appeared elsewhere in the literature. These results are generally applicable to effective BSM theories with B and L violating processes\footnote{It has been noted \cite{Perez:2014fja} that the simultaneous presence of B and L violating operators allow a physical interpretation for the $SU(2)$ vacuum angle.}.

\section*{Acknowledgements}
We thank Chris Brust, John Ellis, Sebastian Ellis, Gordon Kane, David E. Kaplan, Eric Kuflik, Aaron Pierce, Prashant Saraswat, Matthew Walters, and Junjie Zhu for helpful discussions. We are grateful for the excellent atmosphere and facilities at the University of Colorado, Boulder  during TASI 2013, when this collaboration was initiated. We thank the organizers of TASI 2013, in particular Bogdan Dobrescu and Iain Stewart. CF is supported by NSF grant PHY-1214000. SP is supported in part by DOE grant DE-FG-02-92ER40704.  BZ is supported in part by DOE grant DE-SC0007859.

\appendix

\section{Constraints on $\left|\lambda^{\prime \prime}_{ 1 1 m} \lambda_{i j k}\right|$ from 3-body Nucleon Decay}\label{3bodydecay}

The simultaneous presence of $\lambda^{\prime \prime}_{ 1 1 m}$ and $\lambda_{i j k}, \, i \neq k, j \neq k$ can at tree-level lead to 3-body nucleon decay modes $p \rightarrow \nu \nu \ell^+ $, $n \rightarrow \ell^+ \ell^- \nu$, as noted in \cite{Carlson:1995ji}. Because $\lambda^{\prime \prime}_{ 1 1 1} = 0$ due to $SU(3)_C$ invariance, the leading tree level contributions to these 3-body decay modes require some source of quark/squark flavor changing. In this section, we argue that bounds on $\left|\lambda^{\prime \prime}_{1 1 m} \lambda_{i j k} \right|, \, i \neq k, j \neq k$ from 3-body nucleon decay modes are subdominant to the bounds from $N \rightarrow K \ell^+_{i} \ell^-_{j} \nu_k$ discussed in Section \ref{lightflavorconstraints}, due to flavor suppression of the relevant tree-level diagrams.

\begin{figure}[t!]
\center{
\includegraphics[scale=0.8]{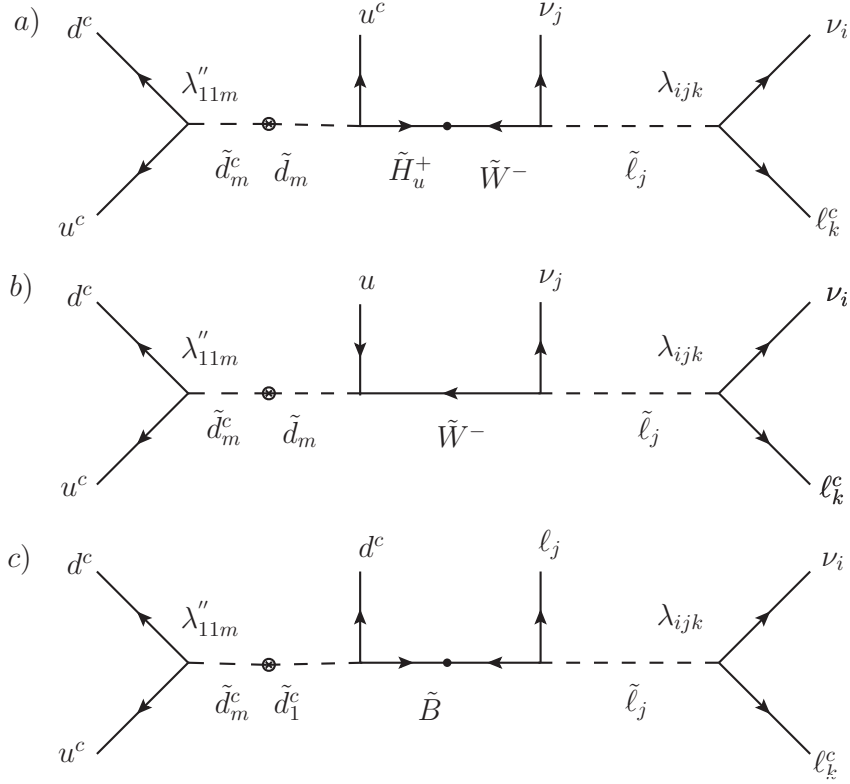}
 \caption{Diagrams involving $\lambda^{\prime \prime}_{1 1 m}$ and $\lambda_{i j k},\, i \neq k, j \neq k$ which result in effective operators that mediate 3-body nucleon decay. Decay modes of the form $p \rightarrow \ell^+_i\ell^+_j \ell^-_k$ are kinematically forbidden, as the antisymmetry condition on $\lambda_{ijk}$ would imply a $\tau$ in the final state.\label{Fig:3body}}}
\end{figure}

The relevant diagrams are shown in Figure \ref{Fig:3body}, giving rise to the following 6-fermion effective operators: \begin{equation}
C_a \left(u^c d^c\right) \left(u^c \nu_j\right) \left(\nu_i \ell^c_k\right), \,C_b \left(u^c d^c\right) \left(u^\dagger \partial \cdot \overline{\sigma} \nu_j\right) \left( \nu_i \ell^c_j\right),\, C_c \left( u^c d^c \right) \left(d^c \ell_j \right) \left(\ell^c_k \nu_i\right).
\end{equation} The effective operators with coefficients $C_a$ and $C_b$ induce the decay mode $p \rightarrow \nu_i \nu_j \ell_k^+$, while the operator with coefficient $C_c$ induces $n \rightarrow \nu_i \ell^+_j \ell^-_k$. Following the procedure outlined in Section \ref{constraints}, it is straightforward to compute the nucleon decay rates induced by $C_a$ and $C_c$:\begin{equation}\label{3bodyappendix1}
\Gamma(p \rightarrow \nu_i \nu_j \ell^+_k)= \frac{\beta^2 \left|C_a\right|^2}{6144 \pi^3} {M_p}^5, \hspace{6mm} \Gamma(n \rightarrow \nu_i \ell^-_j \ell^+_k)= \frac{\beta^2 \left|C_c\right|^2}{6144 \pi^3} {M_n}^5
\end{equation} For the decay rate induced by the derivative interaction corresponding to $C_b$, we instead use an approximate expression obtained by dimensional analysis:\begin{equation}\label{3bodyappendix2}
\Gamma( p \rightarrow \nu_i \nu_j \ell^+_k) \sim \frac{\alpha^2 \left|C_b\right|^2}{6144 \pi^3} {M_p}^7
\end{equation} As discussed in Section \ref{constraints}, $\beta$ and $\alpha$ correspond to hadronic matrix elements, which from lattice calculations are $\alpha \approx \beta \approx 0.0118\, (\mathrm{GeV})^3$\cite{Aoki:2006ib}.

We now compare these 3-body decay widths to the $N \rightarrow K \ell^+_i \ell^-_j \nu_k$ decay width computed in Section \ref{constraints}. Constraints on the partial lifetimes for $p \rightarrow \nu \nu \ell^+$ and $n \rightarrow \mu^+ e^- \nu$ are similar to the constraints on $N \rightarrow \mu^+ \,\mathrm{inclusive}$, i.e. $\tau(p \rightarrow \nu \nu \ell^+), \tau(n \rightarrow \mu^+ e^- \nu) \lesssim 10^{32}$ years \cite{McGrew:1999nd}. Thus if $\Gamma(p \rightarrow \nu \nu \ell^+), \Gamma(n \rightarrow \mu^+ e^- \nu) \ll \Gamma(N \rightarrow K \nu_i \ell^+_j \ell^-_k \nu)$, the constraints from 3-body decay modes are subdominant.
 
Consider Figure 7a; for degenerate SUSY masses, $C_a \sim A\,\lambda^{\prime \prime}_{1 1 \ell} \lambda_{i j k} V^*_{1 \ell}\, g\, y_u\, m_{d_\ell}/{M_{SUSY}}^6$ where $M_{SUSY}$ is the common superpartner mass scale. We have assumed that the left-right squark mixing term $m_{d_\ell} X_{d_\ell}$ is such that $X_{d_\ell} = M_{SUSY}$, and that Wino-Higgsino mixing angles are $\mathcal{O}(1)$.  As discussed in Section \ref{constraints} $A \approx 0.22$ accounts for the renormalization of these operators from $Q \approx M_{SUSY}$ to $Q \approx \Lambda_{QCD}$ \cite{Ellis:1981tv}. Comparing (\ref{3bodyappendix1}) and  (\ref{4-bodydecay})  then gives:\begin{equation}
\frac{\Gamma(p \rightarrow \nu \nu \ell^+)}{\Gamma\left(p \rightarrow K^+ \ell^+ \ell^- \nu \right)} \sim 10^{3}\, \left|V_{1 \ell}\right|^2\, y_u^2\left( \frac{m_{d_\ell}}{M_{SUSY}}\right)^2 \ll 1
\end{equation} for $M_{SUSY} \gtrsim 100$ GeV. Note that we have omitted a diagram analogous to Figure 7a with Higgsino exchange, as it is suppressed with respect to Figure 7a by a lepton Yukawa coupling. 

Applying a similar analysis to Figure 7b gives $C_b \sim A\,\lambda^{\prime \prime}_{1 1 m} \lambda_{i j k} V^*_{1 \ell}\, g^2 m_{d_\ell}/{M_{SUSY}}^7$ and thus:\begin{equation}
\frac{\Gamma(p \rightarrow \nu \nu \ell^+)}{\Gamma\left(p \rightarrow K^+ \ell^+ \ell^- \nu \right)} \sim 10^{3} \left|V_{1 \ell}\right|^2 \left(\frac{M_p m_{d_\ell}}{{M_{SUSY}}^2} \right)^2 \ll 1 \end{equation} for $M_{SUSY} \gtrsim 100$ GeV. Finally, we consider Figure 7c, which gives $C_c \sim A\,\lambda^{\prime \prime}_{1 1 m} \lambda_{i j k} g_Y^2 \left(\delta^{RR}_{d, 1 m}\right)/{M_{SUSY}}^5$ where $\delta^{RR}_{d, 1 m}$ represents the flavor changing squark mass insertion. Thus:\begin{equation}
\frac{\Gamma(p \rightarrow \nu \nu \ell^+)}{\Gamma\left(p \rightarrow K^+ \ell^+ \ell^- \nu \right)} \sim 10^{3} \left(\delta^{RR}_{d, 1 m}\right)^2.\end{equation} Note that FCNC constraints require $\delta^{RR}_{d, 1 m} \lesssim 0.1$ for $\sim 1$ TeV squarks and gluinos \cite{Altmannshofer:2009ne}.

\section{Loop Function $L_{ijk}$ Calculation}\label{loopcalc}
In this appendix, we calculate the loop functions $L_{ijk}$ which determine the coefficients of the effective operators in (\ref{6fermionloop}). This requires computing the loop diagrams in Figure \ref{Fig:loops}. In the following calculations, we ignore $LR$ mixing in the squark sector. 

The diagram in \ref{Fig:loops}a with $W$ exchange gives the amplitude:\begin{align*}\notag
& \mathcal{M}^W_{i j k} =\frac{g^2}{2} V_{1 i} V_{j 1/2}^* \lambda^{\prime \prime}_{j i k} m_{d_i} m_{u_j} Y_{p_2}^\dagger \overline{\sigma}_\nu \sigma_\mu Y_{p_1}^\dagger \int \frac{d^d l}{(2 \pi)^d} \frac{ g^{\mu \nu} - l^\mu l^\nu/M_W^2}{(l^2 - M_W^2)((p_1+l)^2 - m_{d_i}^2)((p_2-l)^2-m_{u_j}^2)}\notag \\ &= i \frac{g^2}{8 \pi^2} V_{1 i} V_{j 1/2}^* \lambda^{\prime \prime}_{j i k} m_{d_i} m_{u_j}  Y_{p_2}^\dagger Y_{p_1}^\dagger\left[C_0(M_W^2,m_{d_i}^2,m_{u_j}^2) -{M_W}^{-2}C_{24}(M_W^2,m_{d_i}^2,m_{u_j}^2)\right]\, + \,....\end{align*} 
where $V_{i j }$ is the CKM matrix, $v = 174$ GeV, and we use 2-component notation \cite{Dreiner:2008tw} for the final state spinors. In the second line we have written the loop integral in terms of the well-known Passarino-Veltman 3-point functions \cite{Passarino:1978jh}, omitting terms proportional to the external momenta. Note that $C_0(m_1^2, m_2^2, m_3^2)$ in our notation corresponds to $C_0(0,0,m_1^2,m_2^2,m_3^2)$ in the notation of  \cite{Passarino:1978jh}; we use a similar notation for $C_{24}$.

The diagram in \ref{Fig:loops}b with charged Higgs exchange gives the amplitude:\begin{align*}
\mathcal{M}_{i j k}^{H} & =  - V_{1 i} V_{j 1/2}^* \frac{m_{d_i} m_{u_j} }{v^2} \lambda^{\prime \prime}_{j i k} Y_{p_2}^\dagger \overline{\sigma}^\mu \sigma^\nu Y_{p_1}^\dagger \int \frac{d^4 l}{2 \pi^d} \frac{(p_2-l)_\mu (p_1 + l)_\nu}{(l^2 - m_{H^+}^2)((p_1+l)^2-m_{d_j}^2)((p_2 - l)^2 - m_{u_i}^2)} \\& = \frac{i}{4 \pi^2} V_{1 i} V_{j 1/2}^* \frac{m_{d_i} m_{u_j} }{v^2} \lambda^{\prime \prime}_{j i k }Y_{p_2}^\dagger Y_{p_1}^\dagger C_{24}\big(M_{H^+}^2,m_{d_i}^2,m_{u_j}^2 \big) \,+\, .... \end{align*} The chargino exchange diagram \ref{Fig:loops}c  (which involves only scalar 3-point integrals) contributes the amplitude:\begin{align*}
\mathcal{M}^{\chi}_{i j k } =  \frac{i}{16 \pi^2}\left(\frac{m_{u_i} m_{d_j}}{v^2}\right)& V_{1 j} V_{i 1/2}^*  m_{\chi_l} Y_{p_2}^\dagger Y_{p_1}^\dagger \Big[-\sqrt{2}M_W\lambda^{\prime \prime}_{i j k} \Big( \frac{\mathrm{U}_{l 2} \mathrm{V}_{l 1}}{\cos\beta} C_0\big(m_{\chi_l}^2, m_{\tilde{d}_{R,i}}^2,  m_{\tilde{u}_{L,j}}^2\big) \\& + \frac{\mathrm{U}_{l 1} \mathrm{V}_{l 2}}{\sin\beta} C_0 \big(m_{\chi_l}^2,  m_{\tilde{d}_{L,i}}^2,  m_{\tilde{u}_{R,j}}^2\big)\Big)+ A^{\prime \prime}_{i j k} \frac{\mathrm{U}_{l 2} \mathrm{V}_{l 2}}{\cos\beta \sin\beta} C_0 \big(m_{\chi_l},  m_{\tilde{d}_{R,i}}^2,  m_{\tilde{u}_{R,j}}^2\big)\Big] \end{align*} where we again omit terms proportional to the external momenta. 
$\mathrm{U}, \mathrm{V}$ are the chargino diagonalization matrices \cite{Haber:1984rc} and $A^{\prime \prime}_{i j k}$ is a soft breaking trilinear defined as $\mathcal{L}_{SOFT} \supset -\frac{1}{2} \lambda^{\prime \prime}_{l m n } A^{\prime \prime}_{i j k} \tilde{u}^c_l \tilde{d}^c_m \tilde{d}^c_n$.

Finally, using the definition $i\lambda^{''}_{ijk}L_{ijk} Y_{p_2}^\dagger Y_{p_1}^\dagger\equiv \mathcal{M}^{H}_{ijk}+\mathcal{M}^{W}_{ijk}+\mathcal{M}^{\chi}_{ijk}$, we obtain:\begin{align}\label{loopfunction}\notag
& L_{i j k} = \left(\frac{m_{u_i} m_{d_j}  V_{1 j} V_{i 1}^* }{16 \pi^2 v^2}\right)  \Bigg[ M_{H^+}^2 C_{0}\big(M_{H^+}^2,m_{d_j}^2, m_{u_i}^2 \big) + 3 M_W^2 C_{0}\big(M_{W}^2,m_{d_j}^2,m_{u_i}^2 \big) \\ & \notag
\hspace{1cm}- \sum_{l = 1}^2 M_{\chi_l} \Bigg(\sqrt{2} M_W \frac{\mathrm{U}_{l 2} \mathrm{V}_{l 1}}{\cos\beta} C_0\big(M_{\chi_l}^2, m_{\tilde{d}_{R,i}}^2,  m_{\tilde{u}_{L,j}}^2\big) + \sqrt{2}M_W\frac{\mathrm{U}_{l 1} \mathrm{V}_{l2}}{\sin\beta} C_0 \big(M_{\chi_l}^2, m_{\tilde{d}_{L,i}}^2,  m_{\tilde{u}_{R,j}}^2\big) \\ & \hspace{7.5cm}- A^{\prime \prime}_{i j k} \frac{\mathrm{U}_{l 2} \mathrm{V}_{l 2}}{\cos\beta \sin\beta} C_0 \big(M_{\chi_l}^2, m_{\tilde{d}_{R,i}}^2,  m_{\tilde{u}_{R,j}}^2\big)\Bigg)\Bigg]
\end{align} This result vanishes in the SUSY limit, as expected. Note the divergent pieces of $\mathcal{M}^W_{i j k }$ and $\mathcal{M}^H_{i j k }$ which are contained within the $C_{24}$ term cancel in $L_{i j k }$. The scalar 3-point function $C_0$ is given by (see e.g. \cite{Logan:1999if}):\begin{equation}
C_0 (m_1^2, m_2^2, m_3^2) = \frac{m_1^2 m_2^2 \log\left(m_1^2/m_2^2\right)+m_2^2 m_3^2 \log\left(m_2^2/m_3^2\right) + m_3^2 m_1^2 \log\left(m_3^2/m_1^2\right)}{(m_1^2-m_2^2)(m_2^2-m_3^2)(m_3^2-m_1^2)}
\end{equation}

\section{A Flavor Model for Right-Handed Neutrinos}\label{flavor}

In this Appendix, we illustrate the constraints which can arise if the dynamics that breaks $Z_2^e \times Z_2^\mu \times Z_2^\tau$ to generate neutrino masses also generates dangerous LRPV operators. For concreteness, we focus on a right-handed neutrino model involving a spurious $SU(3)_\ell \times SU(3)_N$ flavor symmetry. $Z_2^e \times Z_2^\mu \times Z_2^\tau$ can naturally be realized as a sugbroup of $SU(3)_{\ell}$ if both $L$ and $E^c$ transform in the $\overline{{\tiny \yng(1)}}$ representation of $SU(3)_\ell$. We take the right-handed neutrinos $N$ to transform in the ${\tiny \yng(1)}$ representation of $SU(3)_N$. Requiring $SU(3)_\ell \times SU(3)_N$ invariance of the charged lepton and neutrino Yukawa couplings along with the $N$ majorana mass term fixes the transformation properties of the flavor spurions. The matter and representation content of this model is summarized in Table \ref{Table1}. Note that the same $SU(3)_N$ representation content was considered in \cite{Csaki:2011ge}; we will use some results obtained by these authors in the following analysis.

The $SU(3)_\ell \times SU(3)_N$ invariant superpotential to lowest order in spurions is:\begin{equation}\label{flavorsuperpotential}
W = \left(Y_E\right)^{i j} L_{i} E^c_{j}  H_d+ \left(Y_\nu\right)^{i}_a L_i H_u N\hspace{0.5mm}^a+ \left(Y_N\right)_{a b}\Lambda_R  N^a\, N^b+\frac{\lambda}{2} \epsilon^{i j k} L_i L_j E^c_k. \end{equation} Here raised/lowered $i, j, k$ represent fundamental/antifundamental indices transforming under $SU(3)_\ell$, and $a, b, c$ represent fundamental/antifundamental  $SU(3)_N$ indices. We treat $\Lambda_R$ as an undetermined mass scale. $Y_E$ can be thought of as a symmetric matrix which transforms under $SU(3)_\ell$ as $Y_E \rightarrow U^T Y_E U$, so without loss of generality, we take (\ref{flavorsuperpotential}) to be in the basis where $Y_E$ is diagonal. 

The non-zero \emph{vev} of $Y_\nu$ breaks $Z_2^e \times Z_2^\mu \times Z_2^\tau$ and regenerates dangerous $q q q \ell$ operators. To estimate the coefficients of these dangerous operators and the resulting bounds on the neutrino sector, we perform a spurion analysis focusing on $SU(3)_N$ singlets which transform as non-trivial irreps of $SU(3)_\ell$. Let us first consider holomorphic products of $Y_N$, $Y_\nu$. As shown in \cite{Csaki:2011ge}, there are only two irreducible holomorphic $SU(3)_N$ singlets which can be formed from the $Y_\nu$ and $Y_N$ spurions:\begin{equation}\label{holosinglets}
\left(\mathcal{Y}_1\right)_{i j}= \big(\tilde{Y}_\nu\big)^{a}_{i} \big(Y_{N}\big)_{a b} \big(\tilde{Y}_\nu\big)^{b}_j \hspace{10mm} \left(\mathcal{Y}_2\right)^{i j}= \big(Y_\nu\big)_{a}^i \big(\tilde{Y}_N\big)^{a b} \big(Y_\nu\big)_{b}^j
\end{equation} where  $\epsilon_{a c d} \, \tilde{Y}\hspace{0.5mm}^{a b} =\epsilon^{b e f} \,Y_{c e} Y_{d f}$. The non-holomorphic $SU(3)_N$ singlets which transform as non-trivial $SU(3)_\ell$ irreps are given to leading order in $Y_N$, $Y_\nu$ by \cite{Csaki:2011ge}:\begin{equation}\label{nonholosinglets}
\left(\mathcal{V}_1\right)^{i j } \equiv (Y_\nu)^i_a (Y_N^\dagger)^{a b} \left(Y_\nu\right)^j_{b}, \hspace{3mm} \left(\mathcal{V}_2\right)^i = \epsilon^{i  j k} (Y_{\nu}^\dagger)^a_j (Y_N)_{a b} (\tilde{Y}_{\nu})^{b}_k, \hspace{3mm} \left(\mathcal{V}_3\right)^i_j = (Y_\nu^\dagger)^{a}_j (Y_\nu)^i_a
\end{equation} 

\begin{table}
\centering
\begin{tabular}{|c | c | c | c | c| c| c| c | c | c | c | c |}
\hline
 & $L$ & $E^c$ & $N$& $Y_E$ & $Y_\nu$ & $Y_N$ & $\mathcal{Y}_1$ & $\mathcal{Y}_2$ & $\mathcal{V}_1$ & $\mathcal{V}_2$ & $\mathcal{V}_3$\\
\hline
$SU(3)_{\ell}$ & $\overline{\tiny\yng(1)}$ & $\overline{\tiny\yng(1)}$ & 1 & ${\tiny\yng(2)}$ & ${\tiny\yng(1)}$ & 1 & $\overline{\tiny\yng(2)}$  & ${\tiny\yng(2)}$ & ${\tiny\yng(2)}$ & \tiny\yng(1)  & Adj.\\
\hline
$SU(3)_N$ & $1$ & $1$ &${\tiny\yng(1)}$ & $1$ & ${\tiny\overline{\yng(1)}}$ & ${\tiny\overline{\yng(2)}}$ & 1 & 1 & 1 & 1 & 1\\
\hline
\end{tabular}
\caption{Matter and representation content of the $SU(3)_\ell \times SU(3)_N$ model considered in this section. $\mathcal{Y}_1,\, \mathcal{Y}_2$ and $\mathcal{V}_1, \,\mathcal{V}_2, \,\mathcal{V}_3$ are $SU(3)_\ell$ irreps formed from products of the $Y_N, Y_\nu$ spurions defined in (\ref{holosinglets}), (\ref{nonholosinglets}). \label{Table1}}
\end{table}

To obtain bounds on $Y_\nu$ and $Y_N$ from 2-body nucleon decay, we consider $SU(3)_\ell$ singlets formed out of the spurions (\ref{holosinglets})-(\ref{nonholosinglets}) which generate LRPV operators constrained by 2-body nucleon decay\footnote{In performing this analysis, it is useful to keep in mind the $Z_3$ center of $SU(3)_\ell$, under which ${\tiny \yng(1)} \rightarrow e^{2 \pi i/3}\,{\tiny \yng(1)}$ and ${\tiny \yng(2)} \rightarrow e^{4 \pi i/3}\, {\tiny \yng(2)}\,$.}. The leading holomorphic $SU(3)_\ell \times SU(3)_N$ singlet involving a single lepton field is given by:\begin{equation}\label{holospurion}
\mathcal{W}^i L_i \equiv \epsilon_{k m n}\big(\mathcal{Y}_{1}\big)_{p j} \big(Y_E\big)^{k p} \big(Y_E \big)^{m j} \big(Y_E \big)^{n i} L_i \sim  {Y_\nu}^4 Y_N \left(\frac{m_e m_\mu m_\tau}{v^3 \cos^3\beta}\right) L_i
\end{equation}where we have assumed an anarchic flavor structure for $Y_\nu$, $Y_N$. There is also a singlet of the form $(\mathcal{Y}_2)^2 (Y_E)^3 L$, but it is higher order in $Y_N$ compared to (\ref{holospurion}). The leading non-holomorphic singlet involving a single lepton field is given by:\begin{equation}\label{nonholospurion}
\mathcal{K}^i L_i  \equiv \epsilon^{i j k} (Y_E^\dagger)_{j l}\big(\mathcal{V}_3\big)^l_k  L_i \sim {Y_\nu}^2 \left(\frac{m_\tau}{v \cos\beta}\right)L_{1, 2}+ ...
\end{equation} There are similar non-holomorphic terms generated by $\mathcal{V}_1$ and $\mathcal{V}_2$, but they are higher order in $Y_N, Y_\nu$. There are also Kahler potential terms involving non-holomorphic spurions which can induce flavor-changing slepton mass insertions:\begin{equation}
K = \beta_1 {L^\dagger}^i (\mathcal{V}_3)_i^j L_j + \beta_2 {{E^c}^\dagger}^i (\mathcal{V}_3)_i^j {E^c}_j + ...
\end{equation}

Thus to leading order in the spurion analysis, the dangerous $Z_2^e \times Z_2^\mu \times Z_2^\tau$ violating operators are given by:\begin{align}
\notag & W_{spurion} = \alpha_{x y} \mathcal{W}^i \,Q_x L_i  D^c_y + \kappa_0 \mathcal{W}^i L_i H_u + m_{soft} \mathcal{K}^i L_i H_u \\
\label{spurionsuperpotential} &\mathcal{L}_{soft} = \beta_1  {M_{SUSY}}^2 \tilde{\ell}^{\dagger i} (\mathcal{V}_3)_i^j \tilde{\ell}_j + \beta_2 {M_{SUSY}}^2 {\tilde{\ell}}^{c \dagger}\,^i (\mathcal{V}_3)_i^j \tilde{\ell}^c_j  + ...\end{align} where $x, y$ represent arbitrary quark flavor indices. $m_{soft}$ is generated by Planck suppressed Kahler potential operators upon SUSY breaking, i.e. $m_{soft} \approx \left<F_X\right>/M_{pl}$ where $X$ is a SUSY breaking spurion. The non-holomorphic spurions will also generate trilinear RPV terms upon SUSY breaking; we assume that SUSY breaking renders these contributions negligible i.e. $\left<F_{X}\right>/M_{pl}^2 \ll 1$. We now discuss bounds on the neutrino sector in light of (\ref{spurionsuperpotential}) and the 2-body nucleon decay bounds (\ref{2bodyconstraint}). Assuming an anarchic flavor structure for $Y_N$ and $Y_\nu$, the SM neutrino masses are given by $m_\nu \sim {Y_\nu}^2 v^2 \sin^2\beta/M_R$ where $M_R \equiv Y_N \Lambda_R$; we use this relation to fix $Y_\nu$.  We also assume that some mechanism enforces $\kappa_0 \lesssim \mu$. 

Let us first consider contributions from the holomorphic term. Taking $\alpha_{x y} \sim \mathcal{O}(1)$, the holomorphic spurion $\mathcal{W}^i$ generates an effective $Q L D^c$ term\footnote{This spurion can also generate $\lambda_{i k k }L_i L_k E^c_k$ couplings, but bounds on these couplings are weaker than bounds on $\lambda^\prime Q L D^c$; see (\ref{2bodyconstraint}).} with $\lambda^\prime \approx \mathcal{W}^i$. Combining (\ref{holospurion}) and (\ref{spurionsuperpotential}), the constraint on $\lambda^\prime$ in (\ref{2bodyconstraint}) can be mapped into a constraint on $M_R$:\begin{equation}\label{bound1} M_R \lesssim \, 10^{11}\, \mathrm{GeV}\times \left(\frac{10^{-7}}{\lambda^{\prime \prime}_{1 1 2}}\right)^{1/2}\left(\frac{10^{-4}}{Y_N}\right)^{1/2}\left(\frac{10}{\tan\beta}\right)^{3/2}\left(\frac{0.1 \, \mathrm{eV}}{m_{\nu}}\right)\left(\frac{M_{SUSY}}{\mathrm{TeV}}\right)\end{equation} Thus taking typical parameter values and imposing the di-nucleon decay constraint $\left|\lambda^{\prime \prime}_{1 1 2} \right| \lesssim 10^{-6} (M_{SUSY}/\mathrm{TeV})^{5/2}$\cite{Goity:1994dq}), bounds from the holomorphic spurion are rather mild and allow for $M_R$ to lie well above the TeV scale.
 
Let us now consider the non-holomorphic superpotential term, which generates effective $\kappa^i L_i H_u$ couplings with $\kappa^i \approx m_{soft} \mathcal{K}^i$. Combining (\ref{nonholospurion}) and (\ref{spurionsuperpotential}) with (\ref{2bodyconstraint}), the bound on $M_R$ due to $\mathcal{K}^i$ is:\begin{equation} \label{bound2} M_R \lesssim 0.1\, \mathrm{GeV} \times \left(\frac{\mu}{m_{soft}}\right) \left(\frac{10}{\tan\beta}\right)^2 \left(\frac{0.1 \, \mathrm{eV}}{m_\nu}\right) \left(\frac{M_{SUSY}}{\mathrm{TeV}}\right)^2 \end{equation} If $m_{soft} \sim \mu$, this constraint is much stronger than the constraint from the holomorphic spurion, and will not allow $M_R$ to lie above the weak scale. However, the ratio $\mu/m_{soft}$ depends on SUSY breaking dynamics. Taking $\mu \sim M_{SUSY}$ and $m_{soft} \approx \left<F_X \right>/M_{pl} \sim \Lambda_{M} M_{SUSY}/M_{pl}$ where $\Lambda_{M}$ is the mass scale of SUSY breaking messengers, we obtain $\mu/m_{soft} \sim M_{pl}/\Lambda_{M}$. Thus if the messenger scale is comparable to the Planck scale (as in gravity mediation), $M_R$ is constrained to lie below a GeV or so. However, if the messenger scale is hierarchically smaller than $M_{pl}$ (as in gauge mediation), the resulting suppression of $m_{soft}$ can allow $M_R$ to lie above the TeV scale while still satisfying (\ref{bound2}).

Finally, we consider bounds due to flavor-changing slepton mass insertions generated by $\mathcal{V}_3$ which in conjunction with $\lambda^{\prime \prime}_{1 1 2}$ and $\lambda_{i j k }$, $i \neq k, j \neq k$ results in diagrams similar to Figure \ref{Fig:2body}.c. This results in the bound:\begin{equation}\label{sleptonmassinsertion}
M_R \lesssim 10^6\, \mathrm{GeV} \times \left(\frac{10^{-7}}{\lambda^{\prime \prime}_{1 1 2}}\right) \left(\frac{10^{-5}}{\lambda_{i j k}}\right) \left(\frac{0.1 \, \mathrm{eV}}{m_{\nu}}\right) \left(\frac{M_{SUSY}}{\mathrm{TeV}}\right)^3 \frac{1}{\beta_{1, 2}}
\end{equation} Note that neutrino mass constraints require $\left|\lambda_{1 2 3 } \lambda_{1 3 2 } \right| \lesssim 10^{-6}(M_{SUSY}/\mathrm{TeV})$\cite{Hall:1983id,Barbier:2004ez}

Thus if the non-holomorphic contributions proportional to $m_{soft}$ are sufficiently suppressed, $M_R$ can lie well above the TeV scale without violating nucleon decay bounds. The implications of these constraints on RH neutrino parameters begs for a UV completion of the spurion model presented here (see e.g. \cite{Krnjaic:2012aj,Franceschini:2013ne,Csaki:2013we}
). We leave this for future work.

\section{Sneutrino 4-Body BRPV Decay Rate}\label{4bodysneutrinodecay}

\begin{figure}
\includegraphics{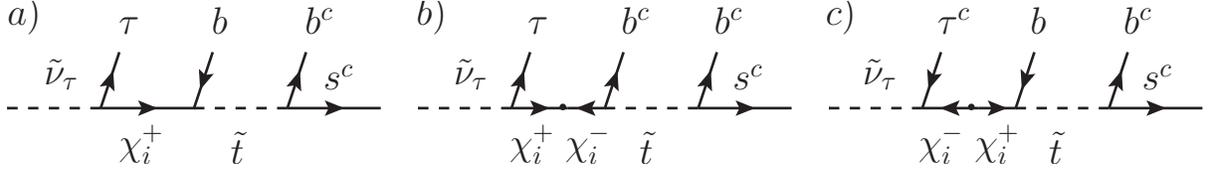}
\caption{Diagrams which contribute to $\tilde{\nu}_\tau \rightarrow \tau \overline{b} \,\overline{b} \,\overline{s}$; we neglect the diagram with pure $\chi^-$ exchange.\label{sneutrino}}
\end{figure}

If BRPV couplings are non-vanishing, a tau sneutrino LSP has the 4-body BRPV decay modes $\tilde{\nu}_\tau\rightarrow \tau \,\overline{d}_i \overline{d}_j \overline{d}_k$ and $\tilde{\nu}_\tau \rightarrow \nu \, u_i d_j d_k$, which occur via virtual neutral/charged-ino and squark exchange. In this appendix, we compute this BRPV decay rate, focusing on the spectra in Figure \ref{spectra} in which $\left|\lambda^{\prime \prime}_{3 3 2}\right| = 0.1$ is the only non-vanishing BRPV coupling. We also assume $m_{\tilde{\nu}_\tau} \lesssim m_t$, such that $\tilde{\nu}_\tau \rightarrow \nu_\tau t b s$ is phase space suppressed. This allows us to consider $\tilde{\nu}_\tau \rightarrow \tau  \overline{b}\, \overline{b} \,\overline{s}$ as the only relevant BRPV decay mode.

The relevant diagrams are depicted in Figure \ref{sneutrino}; we neglect the diagram with pure $\chi^-$ exchange which is proportional to $y_b y_\tau$. The resulting amplitude is given by:
\begin{equation}
i \mathcal{M} = {\lambda^{\prime \prime}_{3 2 3}}^{\hspace{-2mm}*}\, \epsilon_{A B C} \left(\frac{{X_3^B}^\dagger {X_4^C}^\dagger}{{m_{\tilde{\nu}_\tau}}^4}\right) \sum_{a = 1, 2} \sum_{j =1, 2} \left(\frac{c_j^a Y_2^A \left[ - i \left( p - k_1\right) \cdot \sigma \right] X_1^\dagger + m_{\tilde{\nu}_\tau} d_j^a {X_2^A}^\dagger X_1^\dagger+ m_{\tilde{\nu}_\tau} f_j^a Y_2^A Y_1}{ 1- r_{\tilde{t}_a}^2 - z_1 - z_2 + z_{1 2}}\right)
\end{equation} where we have used 2-component spinor notation\cite{Dreiner:2008tw}. We have defined $z_1 = 2 p \cdot k_1/m_{\tilde{\nu}_\tau}^2, \, z_2 = 2 p \cdot k_2/m_{\tilde{\nu}_\tau}^2$,  $z_{12} = 2 k_1 \cdot k_2/m_{\tilde{\nu}_\tau}^2$ and $r_{\tilde{t}_a} = m_{\tilde{t}_a}/m_{\tilde{\nu}_\tau}$; $A, B, C$ denote color indices. Here $p$ is the $\tilde{\nu}_\tau$ 4-momentum, and the label $i$ on the outgoing spinors and momenta $k_i$ increases from left to right with respect to Figure \ref{sneutrino} ($k_1$ is the $\tau$ momenta). The coefficients $c_j^a, \, d_j^a$ and $f_j^a$ correspond respectively to Figure \ref{sneutrino}.a, Figure \ref{sneutrino}.b and Figure \ref{sneutrino}.c:\begin{align}\label{amplitude}
\notag & c_j^a = \frac{g \mathrm{V}_{j 1 } \left( y_t \mathrm{V}_{j 2 }^* R_{t_a}^* -  g \mathrm{V}_{j 1 }^* L_{t_a}^*\right) R_{t_a}}{{r^2_{\chi_j}} - 1 + z_1} \hspace{1cm} d_j^a = \frac{g \mathrm{V}_{j 1} y_b \mathrm{U}_{j 2} L_{t_a}^* R_{t_a} r_{\chi_j}}{{r^2_{\chi_j}} - 1 + z_1}\\ & \, \hspace{2.5cm} f_j^a = \frac{y_\tau \mathrm{U}_{j 2}^* \left( y_t \mathrm{V}_{j 2 }^* R_{t_a}^* - g \mathrm{V}_{j 1 }^* L_{t_a}^*\right) R_{t_a} r_{\chi_j}}{{r^2_{\chi_j}} - 1 + z_1} \end{align} Note that $y_t, y_b, y_\tau$ are supersymmetric Yukawa couplings i.e. $y_t = m_t /v \sin \beta, \, y_b = m_b/v \cos\beta$. We have defined $r_{\chi_j} = m_{\chi^\pm_j}/m_{\tilde{\nu}_\tau}$. $\mathrm{U}, \mathrm{V}$ are the chargino diagonalization matrices \cite{Haber:1984rc}, and $R_t$, $L_t$ are stop mixing angles defined by:\begin{equation}
\begin{pmatrix}
\tilde{t}_R \\
\tilde{t}_L 
\end{pmatrix} = 
\begin{pmatrix}
R_{t_1} & R_{t_2} \\
L_{t_1} & L_{t_2} \\
\end{pmatrix} \begin{pmatrix}
\tilde{t}_1 \\
\tilde{t}_2
\end{pmatrix}
\end{equation}

In computing the decay rate, we neglect final state fermion masses, allowing us to neglect interference terms between the diagrams in Figure \ref{sneutrino}. The decay rate is then given by:\begin{align}\label{decayrate}
&\notag \Gamma(\tilde{\nu}_\tau \rightarrow \tau \, \overline{b} \, \overline{b}\, \overline{s}) = \int d\Phi_4 \frac{\sum_{spins} \left|\mathcal{M}\right|^2}{2\, m_{\tilde{\nu}_\tau}}\\ & =\int \hspace{-2mm} d\Phi_4 \left(\frac{3 \left|\lambda^{\prime \prime}_{ 3 2 3} \right|^2\left(1 - z_1 - z_2 + z_{1 2}\right)}{{m_{\tilde{\nu}_\tau}}^3}\right)  \sum_{a, b} \sum_{j, k} \frac{c_j^a {c_k^b}^* \left(z_1 z_2 - z_{1 2}\right) + \left(d_j^a {d_k^b}^*  + f_j^a {f_k^b}^* \right) z_{1 2}}{\left(1 - r^2_{\tilde{t}_a} - z_1 - z_2 + z_{1 2} \right)\left(1 - r^2_{\tilde{t}_b} - z_1 - z_2 + z_{1 2} \right)}
\end{align} where the sums on $a, b, j, k$ run from 1 to 2. We perform the 4-body phase space integration using the RAMBO method \cite{Kleiss:1985gy}, which is particularly straightforward to implement assuming massless final state particles.

The result for $m_{\tilde{\nu}_\tau} = 100$ GeV is shown in Figure \ref{sneutrinoplot}, which plots contours of constant $c \tau(\tilde{\nu}_\tau \rightarrow \tau \overline{b}\, \overline{b} \,\overline{s})$ in the $\left(\mu, m_{\tilde{t}}\right)$ plane where we take $m_{\tilde{t}}$ to be the degenerate stop mass. The left (right) plot is for $\tan \beta = 2\, ( 10) $;  $M_2 = 300$ GeV and vanishing stop mixing is assumed for both plots. Note from (\ref{amplitude}) that in the limit of vanishing stop mixing and vanishing Wino-Higgsino mixing, $c_j^a = 0, d_j^a = 0$, and only the pure Higgsino portion of $f_j^a$ is non-vanishing. Thus for $M_2, \mu \gg M_Z$, the $M_2$ dependence is weak while the $\tan \beta$ dependence is strong, as evident in Figure \ref{sneutrinoplot}.

\bibliographystyle{utcaps}
\bibliography{bib}

\end{document}